\documentclass[review,sort&compress]{elsarticle}
\usepackage[utf8]{inputenc}
\usepackage[monochrome]{xcolor}
\usepackage{fullpage} 
\usepackage{hyperref}
\usepackage{multirow}
\usepackage{lineno}
\usepackage{enumerate}
\usepackage{tikz}
\usepackage{url}
\modulolinenumbers[5]
\usepackage{stmaryrd}

\usepackage{nomencl}
\usepackage{amsmath,mathtools}
\usepackage[capitalise, noabbrev, nameinlink]{cleveref}
\usepackage{amsfonts,bm}
\usepackage{multirow}
\usepackage{subfig}

\usepackage{graphicx}
\usepackage{booktabs}
\usepackage{makecell}
\usepackage{siunitx}
\usepackage[skip=0.6\baselineskip]{caption}

\usepackage{xpatch}
\let\oldequation\equation
\let\oldendequation\endequation
\renewenvironment{equation}
  {\linenomathNonumbers\oldequation}
  {\oldendequation\endlinenomath}
  
\renewcommand\nomgroup[1]{
  \item[\itshape
  \ifstrequal{#1}{A}{Symbols}{
  \ifstrequal{#1}{B}{Roman Letters}{
  \ifstrequal{#1}{C}{Greek Letters}{
  \ifstrequal{#1}{D}{Abbreviations}{}}}}
]}

\usepackage{floatrow}
\usepackage[commentmarkup=footnote]{changes}

\usepackage{bm}
\usepackage{todonotes}
\usepackage{url}

\usepackage{longtable,tabularx}

\graphicspath{{./figs/}}

\usepackage{array}
\newcolumntype{P}[1]{>{\centering\arraybackslash}m{#1}}

\begin{document}
\begin{frontmatter}
  
\title{A PDE-free, neural network-based eddy viscosity model coupled with RANS equations}
    \author[tud]{Ruiying Xu}
    \author[vt]{Xu-Hui Zhou}
    \author[pu]{Jiequn Han}
    \author[tud]{Richard P. Dwight}
    \author[vt]{Heng Xiao\corref{cor}}
    \ead{hengxiao@vt.edu}
    \cortext[cor]{Corresponding author}
  
  \address[tud]{Faculty of Aerospace Engineering, Delft University of Technology, Delft, The Netherlands}
  \address[vt]{Kevin T. Crofton Department of Aerospace and Ocean Engineering, Virginia Tech, Blacksburg, Virginia, USA}
  \address[pu]{Center for Computational Mathematics, Flatiron Institute, New York, USA}

\begin{abstract}
Most turbulence models used in Reynolds-averaged Navier-Stokes (RANS) simulations are partial differential equations (PDE) that describe the transport of turbulent quantities. Such quantities include turbulent kinetic energy for eddy viscosity models and the Reynolds stress tensor (or its anisotropy) in differential stress models. However, such models all have limitations in their robustness and accuracy. Inspired by the successes of machine learning in other scientific fields, researchers have developed data-driven turbulence models. Recently, a nonlocal vector-cloud neural network with embedded invariance was proposed, with its capability demonstrated in emulating passive tracer transport in laminar flows. Building upon this success, we use nonlocal neural network mapping to model the transport physics in the $k$--$\varepsilon$ model and couple it to RANS solvers, leading to a PDE-free eddy-viscosity model. We demonstrate the robustness and stability of the RANS solver with a neural network-based turbulence model on flows over periodic hills of parameterized geometries. Our work serves as a proof of concept for using a vector-cloud neural network as an alternative to traditional turbulence models in coupled RANS simulations. The success of the coupling paves the way for neural network-based emulation of Reynolds stress transport models.
\end{abstract}
 
  
\end{frontmatter}

\section{Introduction}
Turbulence is a physical phenomenon that is ubiquitous in natural and industrial flows. It is prevalent no matter in internal flows such as fluid inside pipes and channels or exterior flows like the air surrounding airplanes and vehicles. Turbulent flows are characterized by their chaotic nature and the wide range of length and time scales~\cite{pope2001turbulent}, which poses a great challenge in understanding and predicting them. Modelling turbulent flows has thus been an important research topic over the past half-century.

The exact description of fluid flows is given by the Navier-Stokes equations when the continuum assumption applies. However, the computational cost of directly solving the Navier-Stokes equations (DNS) is prohibitively high as it grows cubically ($Re^{3}$) with regards to Reynolds number~\cite{moin1998direct}. Simulating turbulent flows at tractable costs requires additional modelling efforts. The most commonly used turbulent flow simulation techniques are Reynolds-averaged Navier-Stokes (RANS) models. The idea is to decompose the turbulent flow into the mean flow field described by the RANS equations (primary equations) and the fluctuating velocity field whose influence on the mean flow field is modelled by turbulence closures. The construction of closure models usually involves a combination of physical understanding and a calibration procedure accommodating simple canonical flows. This procedure, however, is limited by the capability of incorporating various flow configurations~\cite{xiao2019quantification}. Recent development in machine learning techniques has opened up new avenues for solving the long-standing closure problems~\cite{duraisamy2019turbulence}. Thanks to the high expressive power of machine learning models such as neural networks, it is possible to make use of much larger datasets and obtain more generalizable and/or accurate turbulence models~\cite{ling2016reynolds, wu2018physics, weatheritt2016novel, schmelzer2020discovery}.

\subsection{Invariances in machine-learned turbulence models}\label{directions}

It has been observed that the prediction performance of machine-learned turbulence models can be significantly improved by embedding the known physics~\cite{ling2016machine, kaandorp2020data, frezat2021physical}. Most works along this line aim at embedding the symmetry of the turbulence flows, which refers to the invariance properties of the flow under transformations of the coordinate system. Specifically, the invariance properties include invariance under translation, rotation, and uniform motion (Galilean invariance) of the reference frame. Another property of turbulence flows that is seldom incorporated in data-driven turbulence models is its nonlocality. In most data-driven models, the inputs are only local flow variables (e.g., velocity gradient and its linear combinations) at the point of interest~\cite{wang2017physics, alonso2015machine, kaandorp2020data}. This simplification is only reasonable under the assumption that there is a local balance between the production, redistribution and the dissipation of the Reynolds stress ~\cite{pope2001turbulent}. However, as shown in transport equations of Reynolds stress and other turbulence quantities, turbulence quantities at one point are not only determined by the local flow field; instead, they are greatly influenced by the upstream flow structure and boundary conditions~\cite{gatski1996simulation}.

A vector-cloud neural network (VCNN) has been recently proposed~\cite{zhou2022frame} to incorporate the nonlocality. The idea of the VCNN framework is to use the information from a set of points surrounding the point of interest as the input. One challenge for using a group instead of a single point is to guarantee permutation invariance~\cite{han2017deep}. This invariance comes from the fact that the elements in the input set have no intrinsic ordering, and thus the outputs should depend on the set as a whole but not on the specific ordering of the elements. The permutation invariance brings another challenge in addition to the invariance properties required by the flow physics. 

\begin{figure}[!htb]
\centering
\includegraphics[width=0.96\textwidth]{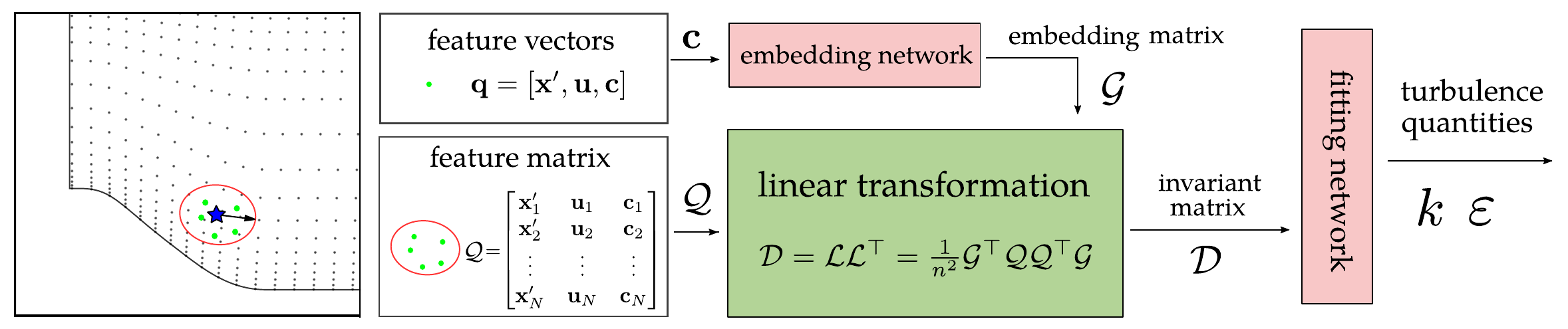}
\caption{Diagram of the VCNN framework for turbulence closure. A stencil near the downhill is visualized as an example.
}
\label{fig:simple_network_diagram}
\end{figure}

In this work, we drew inspirations from the VCNN and developed the neural network shown in \autoref{fig:simple_network_diagram} to emulate the $k$--$\varepsilon$ turbulence model, ensuring all the invariance properties are embedded. The network aims to establish a mapping from mean flow properties in the neighbourhood to turbulence quantities at the point of interest. The network consists of an embedding network and a fitting network, and they are connected by a linear transformation. In the embedding network, higher-level representations of the input feature set are extracted by embedding functions. The linear transformation between two networks guarantees the invariance properties. Finally, the fitting network takes in the invariant feature matrix and provides the final predictions of the targeted variables. Details of the methodology and proof of invariance properties are provided in Section~\ref{methodology_paper}.

The advantages of the proposed framework are its nonlocality, strict invariance-preserving properties and good scalability~\cite{zafar2021frame}. Most of the current machine-learning-based turbulence models are local as mentioned earlier. In comparison, our model takes in information from the neighbouring region and is able to capture more physics in the real flows. Compared to the nonlocal models obtained with other approaches, such as Convolutional Neural Network (CNN)~\cite{guastoni2020prediction, lapeyre2019training, zhou2021learning, gin2020deepgreen}, the proposed model strictly ensures invariances. One of the major drawbacks of CNN is a lack of geometric invariance~\cite{azulay2018deep}. General practice for reducing CNN's deviation of invariance is augmenting the dataset~\cite{shorten2019survey}, in which the input data is transformed by translation and rotation. The enlarged dataset is then fed into the network such that the network acquires certain robustness towards variations of the input data under different coordinate systems. However, as implied by the procedure, there is no guarantee of the exact invariance properties of the model. Minor violation of the symmetries may not influence the classification problem of image recognition, for which CNN was initially developed~\cite{fukushima1988neocognitron}. Nevertheless, in a physics modelling scenario, it may lead to intrinsically wrong results or even breakdowns of the simulation~\cite{han2021machine,zafar2021frame}. Considering the stringent requirement of physics modelling, having solid invariance properties is a major advantage of our framework. Besides, CNN is rarely employed to deal with data on nonuniform or non-Cartesian mesh, which are usually used in fluid simulations for local refinement or adapting complex geometries.

Whilst the previous work on VCNN successfully embedded the nonlocality and frame-independency into the network, there are limitations on the architecture and the application scenario. One line of development is to modify the architecture such that it can be compatible with tensor-based outputs. This is crucial in eventually using VCNN to emulate Reynolds stress transport equations. Such adaptation for equivariant tensor outputs of the network has already been addressed in a follow-up work~\cite{han2022vcnn}. Another line of development stems from the limitation that the performance of the network was only evaluated in predicting the transport of passive scalar in a laminar flow. Its performance in predicting turbulent flow fields is unknown considering the complexity of turbulent flows compared to laminar flows. Furthermore, the final goal of a turbulence model is to serve as a part of the fluid solver and provide stable predictions to the RANS equation solver. Therefore, the neural network model needs to be evaluated in a coupled setup with the RANS solver. This paper aims to address these critical issues.

\subsection{Major challenges in coupling neural network model to RANS Solvers}

Compared to the previous work, the current work of predicting turbulence quantities in a coupled setup is more challenging, due to the interaction between the neural network predictions and the RANS equations. The lack of stability is a common potential problem for neural network models. When coupled with the RANS equations and evaluated in the iteration process of the solver, the neural network model must deal with non-converged flow fields most of the time. Such inputs of mean fields typically are not present in the training data. Consequently, the neural network is required to perform extrapolation tasks during the iterations. 
The behaviour of the networks in such circumstances can be unpredictable. When the predictions provided by the neural network are fed back to the RANS equations, the instability is transported to RANS equations. Studies have shown that the RANS equation can be very sensitive to the Reynolds stresses~\cite{wu2019reynolds}, which poses an additional challenge to the closure modelling. \textcolor{brown}{An alternative strategy for improving the robustness is to incorporate the RANS solver in the training process, but that would introduce major challenges in the training process by requiring adjoint solver~\cite{strofer2021end} or ensemble simulations~\cite{strofer2021ensemble}.}

Apart from the difficulty of extrapolation, the turbulence quantities are more complicated than the concentration field in terms of flow physics. The turbulence quantities are described by two partial differential equations (PDEs) coupled with the source term. In comparison, the concentration is described by a single convection-diffusion equation. In terms of their distribution, the range of magnitude of the turbulence quantities is much larger than that of the concentration field, which makes them numerically more difficult to deal with. Besides, the mean flow field is much more complex with a higher Reynolds number. The existence of a thinner boundary layer indicates more rapidly varying velocities spatially, which can also bring challenges to the neural network model.

\subsection{Contribution of present work}
The present work has two major contributions. First, it extends the previous work on vector-cloud neural networks~\cite{zhou2022frame} from laminar flows to turbulent flows. Instead of predicting the hypothetical passive concentration field, this work directly emulates the transport equations for turbulent kinetic energy ($k$) and dissipation rate ($\varepsilon$), which are of physical significance in turbulence modelling. From the perspective of machine learning, the distributions of turbulence quantities are more challenging due to their large range (for example, in the cases considered in this work, the dissipation rate $\varepsilon$ ranges from $10^{-4}$ to $2$). Special transformation and normalization of the output variables were applied to resolve such a problem. Second, we evaluate the trained neural network model in a coupled setup, which is rarely studied in previous works. Specifically, we couple the neural network-emulated turbulence model with the RANS equation solver to investigate its robustness and stability. 

In this work, the neural network-based turbulence model is trained using the data generated from the $k$--$\varepsilon$ model~\cite{launder1983numerical} and is subsequently evaluated against it. Its accuracy and applicability are thus restricted by the training set itself. We emphasize that the objective here is neither to replace nor to surpass the $k$--$\varepsilon$ model with neural networks. Rather, this work serves as a proof of concept regarding the applicability of the VCNN to turbulence modelling problems. By combining present work with the VCNN architecture with equivariance~\cite{han2022vcnn}, the framework can be further extended to predict Reynolds stress tensor anisotropy. Potential advantages of the integrated framework over traditional models are, for instance, the flexibility in incorporating training data, and less burden of modelling unclosed terms such as the pressure-strain-rate tensor.

The remaining sections of the paper are organized as follows. Section~\ref{methodology_paper} describes the problem to be solved and the methodology used, including the selected input data matrix, the neural network architecture, and the coupling procedure. Section~\ref{results} presents the main results obtained in numerical simulations on periodic hill flow cases with discussions. Section~\ref{conclusion} concludes the paper.

\section{Problem statement and methodology}
\label{methodology_paper}
\subsection{Problem statement}
In RANS simulations, the flow field is described by the mean flow equations and a closure model. For incompressible flows of constant density $\rho$, the mean flow equation is as follows:
\begin{equation}\label{eq:RANS}
    \frac{\bar{\textrm{D}}\langle U_j\rangle}{\bar{\textrm{D}}t}=\nu\nabla^2\langle U_j\rangle-\frac{\partial\langle u_i' u_j'\rangle}{\partial x_i}-\frac{1}{\rho}\frac{\partial \langle p\rangle}{\partial x_j},
\end{equation}
in which $\frac{\bar{\textrm{D}}}{\bar{\textrm{D}}t}\coloneqq\frac{\partial}{\partial t}+\langle \boldsymbol{U}\rangle\cdot\nabla$ is the mean substantial derivative, $\langle U_j\rangle$ is the mean velocity, $u_i'$ and $u_j'$ are the fluctuating velocities, $\langle p\rangle$ is the mean pressure, and $\nu$ is the kinematic viscosity. The unclosed term in Eq.~\eqref{eq:RANS} $\langle u_i' u_j'\rangle$ (the covariance of fluctuating velocities) is also referred to as the Reynolds stress tensor. In the $k$--$\varepsilon$ turbulence model, the unclosed Reynolds stress tensor follows the Boussinesq hypothesis:
\begin{equation}\label{Boussi}
    \langle u_i' u_j' \rangle = \frac{2}{3}k\delta_{ij}-\nu_T\left(\frac{\partial\langle U_i\rangle}{\partial x_j}+\frac{\partial\langle U_j\rangle}{\partial x_i}\right),
\end{equation}
where $\delta_{ij}$ is the Kronecker delta and $\nu_T$ is the eddy viscosity. The eddy viscosity is defined by
\begin{equation}\label{nut_ke}
    \nu_T=C_D\frac{k^2}{\varepsilon},
\end{equation}
where $C_D=0.09$ is a model constant.
The turbulent kinetic energy $k$ and turbulent kinetic energy dissipation rate $\varepsilon$ are described by the transport equations:
\begin{align}
    \frac{\bar{\textrm{D}}k}{\bar{\textrm{D}}t}&=\nabla\cdot\left(\frac{\nu_T}{\sigma_k}\nabla k\right)+\mathcal{P}-\varepsilon\label{transport_equation_k},\\
    \frac{\bar{\textrm{D}}\varepsilon}{\bar{\textrm{D}}t}&=\nabla\cdot\left(\frac{\nu_T}{\sigma_\varepsilon}\nabla\varepsilon\right)+C_{\varepsilon1}\frac{\mathcal{P}\varepsilon}{k}-C_{\varepsilon2}\frac{\varepsilon^2}{k},\label{transport_equation_ep}
\end{align}
in which $\sigma_k=1.00$, $\sigma_\varepsilon=1.30$, $C_{\varepsilon1}=1.44$, and $C_{\varepsilon2}=1.92$, and $\mathcal{P}=2\nu_TS_{ij}S_{ij}$ is the production of $k$ with $S_{ij}$ the strain rate tensor. 

In this work, the goal is to develop a neural network-based turbulence model that plays the role of transport equations~\eqref{transport_equation_k} and~\eqref{transport_equation_ep} in the RANS simulation. The model is nonlocal in the sense that it establishes a mapping from the features of a neighbouring point set (referred to as the stencil or the cloud) to the turbulence quantities $k$ and $\varepsilon$ of the point of interest (referred to as the central point). Meanwhile, the eddy viscosity and Reynolds stress are computed following the same expressions Eq.~\eqref{Boussi} and Eq.~\eqref{nut_ke} as in the $k$--$\varepsilon$ model. Therefore, as with the linear or nonlinear eddy viscosity models, the weak equilibrium assumption is also made in the neural network-emulated $k$--$\varepsilon$ model~\cite{pope2001turbulent, gatski1993explicit}, since only the non-equilibrium in the magnitude of the Reynolds stress is accounted for, and the Reynolds stress anisotropy is still modelled based on local strain rate as shown in Eq.~\eqref{Boussi}.

\subsection{Data generation and processing}\label{data_processing}
The generation and processing of data follow four steps: 
\begin{enumerate}[(1)]
    \item perform simulations with the RANS solver with the $k$--$\varepsilon$ turbulence model,
    \item determine the neighbourhood on which the central point is dependent,
    \item calculate and normalize features and labels,
    \item stack stencils for all the sample points for training or testing the neural network.
\end{enumerate}
They are described in detail in the rest of the section.
\subsubsection{Flow simulation}
\begin{figure}[!htb]
\centering
\includegraphics[width=0.7\textwidth]{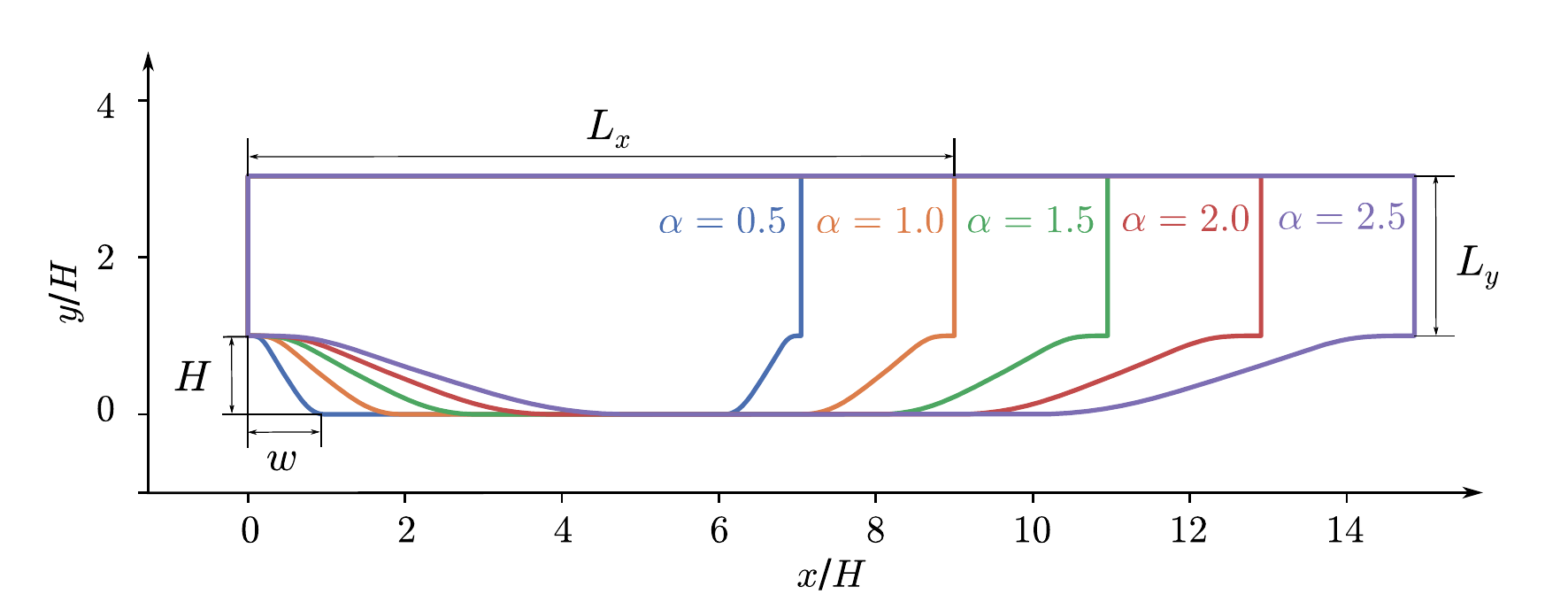}
\caption{Periodic hill geometries with varying slope parameters $\alpha$.}
\label{fig:geometry}
\end{figure}

The flow is simulated in a channel with periodically appearing hills on the bottom~\cite{breuer2009flow}. A family of such periodic hill geometries are parameterized by a slop parameter $\alpha$ as shown in \autoref{fig:geometry}~\cite{xiao2020flows}. The slope parameter $\alpha$ is defined as $\frac{w}{1.93H}$ where $H$ is the height of the hills and $w$ is the width of the hills. It is clear that the smaller the slope parameter, the steeper the bottom of hills. $L_y$ remains the same while $L_x$ is larger for larger slope parameters. A fragment starting with a downslope following a flat region and ending with an upslope is extracted from the periodically varying geometry as the simulation domain. The inlet and outlet are considered as periodic boundaries and the top and bottom of the channel are non-slip walls. 

\begin{figure}[!htb]
\centering
{\hspace{3pt}\includegraphics[height=0.06\textwidth]{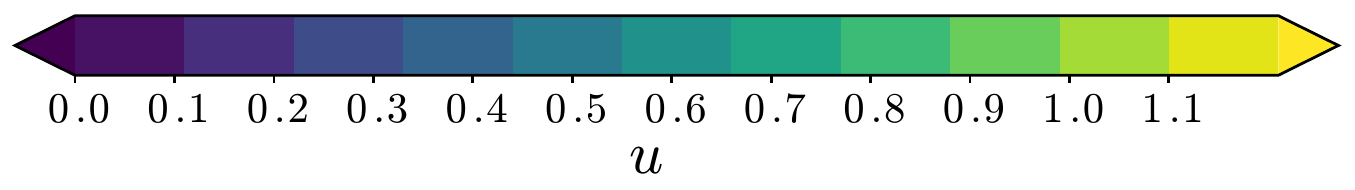}\hspace{15pt}\vspace{5pt}}
\subfloat[Velocity $u$, $\alpha = 1.0$]
{\label{fig:1.0_velocity_contour}\includegraphics[height=0.15\textwidth]{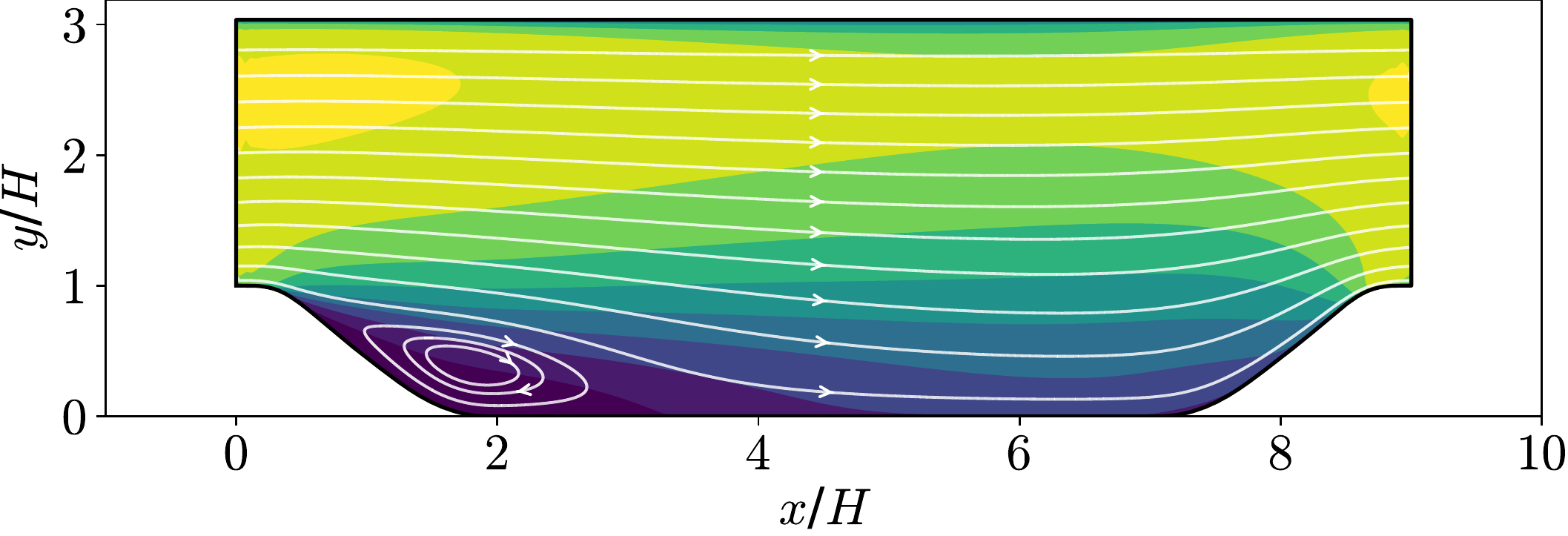}\hspace{1pt}}
\subfloat[Velocity, $\alpha = 2.0$]
{\label{fig:2.0_velocity_contour}\includegraphics[height=0.15\textwidth]{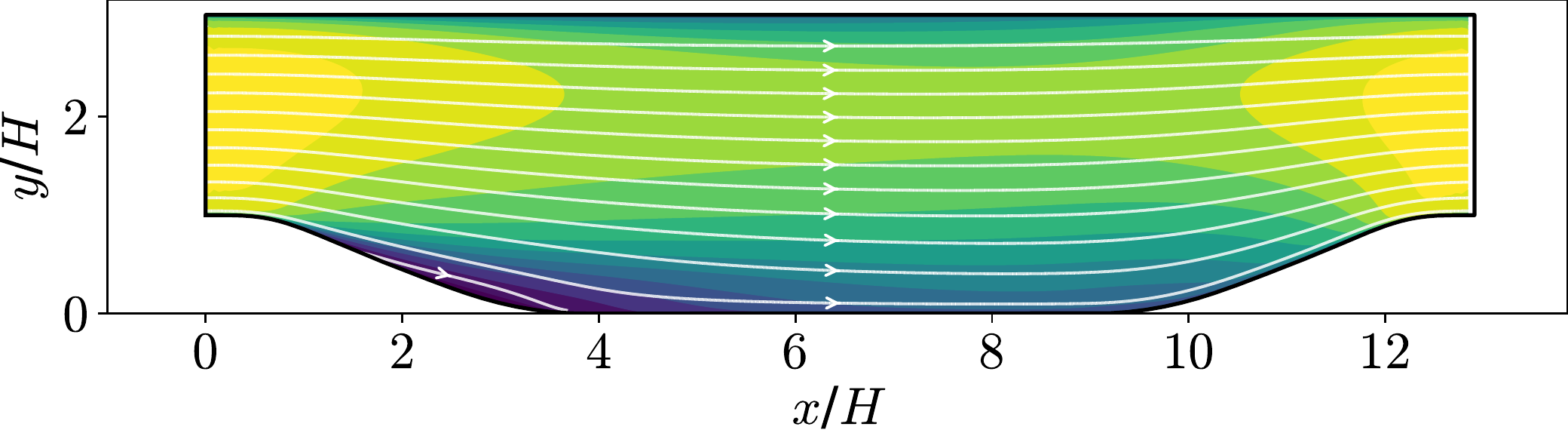}}\\
{\hspace{3pt}\includegraphics[height=0.06\textwidth]{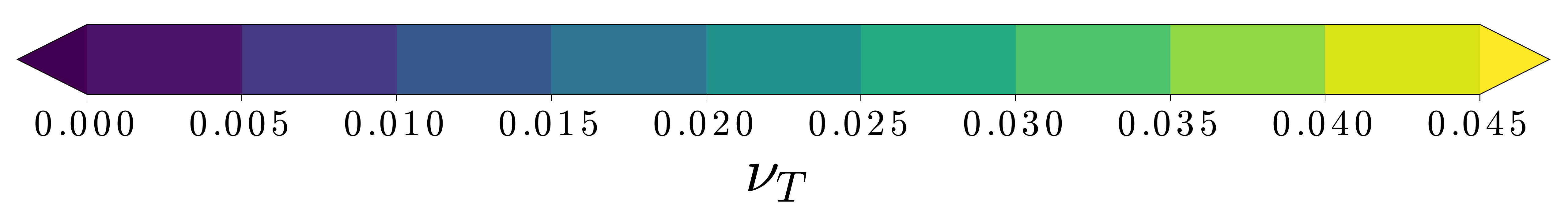}\hspace{15pt}\vspace{0pt}}
{\includegraphics[height=0.06\textwidth]{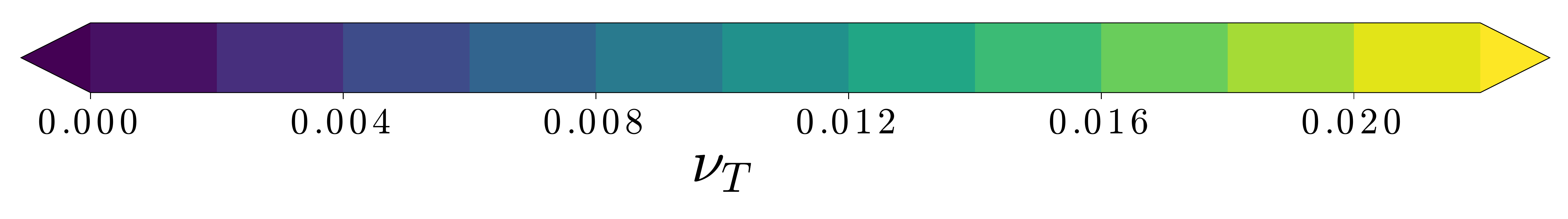}\hspace{15pt}\vspace{0pt}}\\
\subfloat[Eddy viscosity $\nu_T$, $\alpha = 1.0$]
{\label{fig:1.0_nut_contour}\includegraphics[height=0.15\textwidth]{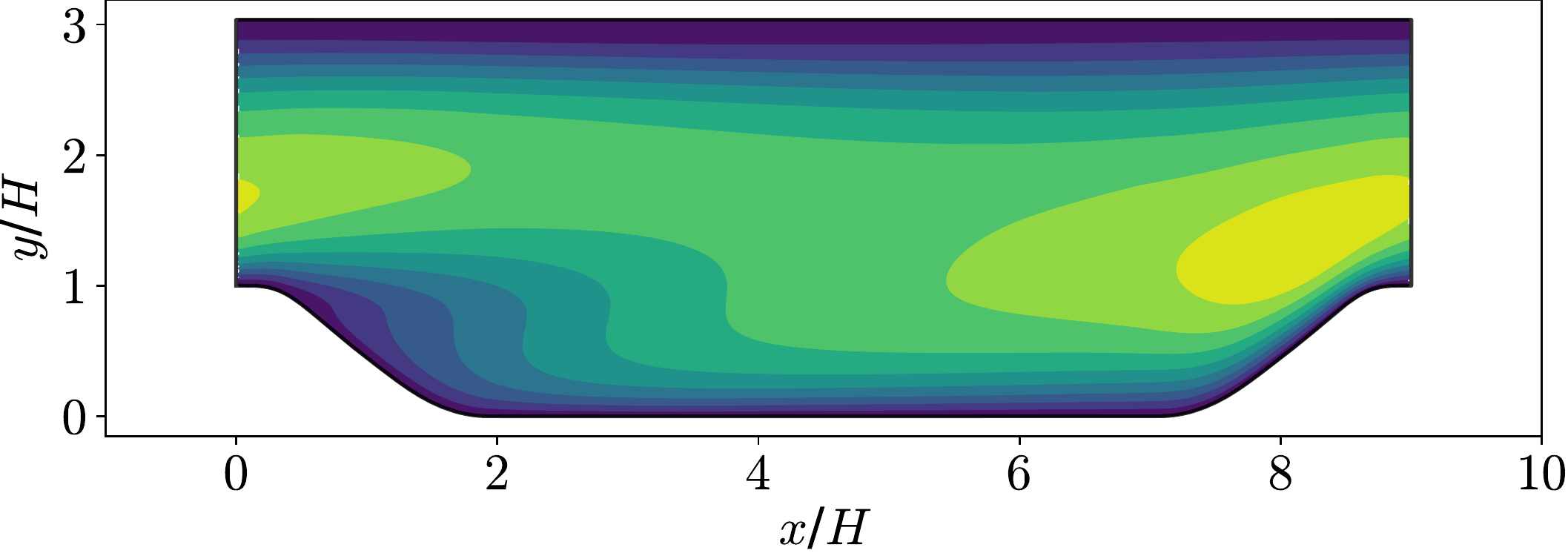}}
\subfloat[Eddy viscosity $\nu_T$, $\alpha = 2.0$]
{\label{fig:2.0_nut_contour}\includegraphics[height=0.15\textwidth]{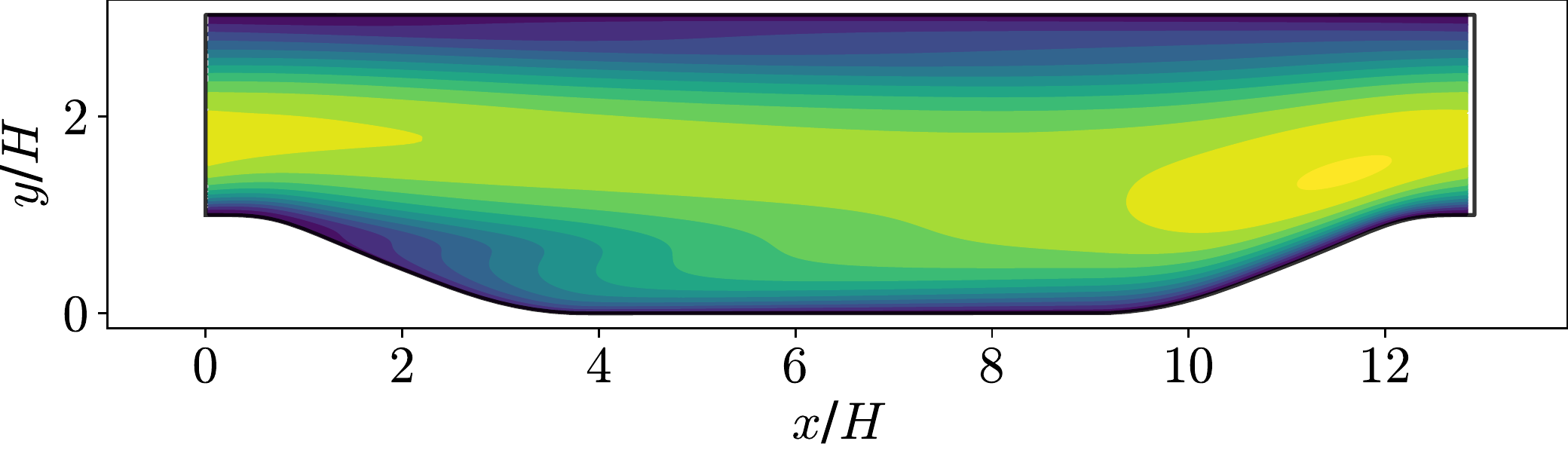}}\\
\caption{Representative simulation results. Panels (a) and (b) present the mean velocity contour and the streamline of the flows in the cases of $\alpha=1.0$ and $\alpha=2.0$. Panels (c) and (d) show the eddy viscosity contour of the respective flow cases.}
\label{fig:simulation_results}
\end{figure}

The flow domain is discretized into a structured mesh with the number of cells in $y$-direction $N_y = 200$ and \textcolor{brown}{the number of cells in $x$-direction increases as the slope parameter $\alpha$ increases.} For instance, the number of cells for case $\alpha=1$ is 40000 and that for $\alpha=1.5$ is 44000. The flow Reynolds number based on the hill-height is kept at 10595 by adjusting the pressure gradient for different slopes. The CFD simulations are performed with the open-source software package OpenFOAM~\cite{weller1998tensorial}. The SIMPLE algorithm~\cite{patankar1983calculation} is adopted for solving the momentum and pressure equations of the incompressible flow. Some of the representative flow fields are visualized in \autoref{fig:simulation_results}. 
The flow case $\alpha=1$ has a larger velocity overall. According to the streamline, the recirculation zone of the flow case $\alpha=1$ can be identified easily. In comparison, in the flow case $\alpha=2.0$, the recirculation is barely visible. The recirculations are more prominent in the geometries with smaller slope parameters. In terms of the eddy viscosity, the flow case $\alpha=1.0$ is characterized by a higher level of eddy viscosity. The distribution patterns of both cases are quite similar.

A group of 11 flow cases with slope parameters $\alpha$ between 1 and 2 are selected to construct the training dataset. They are distributed linearly as $\alpha=1.0, 1.1, 1.2,...,$ and $2.0$. The number of mesh points of each case varies from 40000 ($\alpha=1.0$) to 48000 ($\alpha=2.0$). There are 484000 mesh points in the whole training set and each has 200 samples in their stencil with 11 features for each sample. Outputs for each point are turbulence kinetic energy $k$ and turbulence kinetic energy dissipation rate $\varepsilon$.

The testing dataset comprises 10 extrapolation flow cases and 10 interpolation flow cases. Half of the extrapolation cases have slope parameters larger than 2 and another half have slope parameters smaller than 1. The interpolation test cases are between two neighbouring training cases. The composition of the training and testing set are summarized in \autoref{tab:datasets}.
\begin{table}[H]
    \centering
    \begin{tabular}{llc}
    \Xhline{4\arrayrulewidth}
         Dataset&slope parameters $\alpha$& number of cases\\
         \hline
         Training set &1.0, 1.1, 1.2, 1.3, $\ldots$, 1.8, 1.9, 2.0 &11\\
         Interpolation testing set &1.05, 1.15, 1.25, $\ldots$, 1.85, 1.95&10\\
         Extrapolation testing set &0.9, 0.8, 0.7, 0.6, 0.5, 2.1, 2.2, 2.3, 2.4, 2.5 &10\\
    \Xhline{4\arrayrulewidth}
    \end{tabular}
    \caption{Datasets for training and testing}
    \label{tab:datasets}
\end{table}

\subsubsection{Influence region}

According to the nonlocality of the turbulence physics and the nature of the transport equations for $k$ and $\varepsilon$, each point in the flow is affected by the entire history of the fluid particle, as well as pressure-coupling with other regions. The VCNN incorporates part of the nonlocality by considering the stencil containing the neighbouring points of the point of interest. It is presumed that the stencil has the shape of an ellipse with its center located at the central point and its major axis aligned to the mean velocity at that point. The idea is that the diffusion is isotropic whilst the advection is strongest along the local flow velocity direction. The lengths of the semi-major and semi-minor axes are determined according to the following equations~\cite{zhou2021learning}:
\begin{align}
    l_1=\left|\frac{2\nu_l\log\epsilon}{\sqrt{|\boldsymbol{u}_a|^2+4\nu_l\zeta_l}-|\boldsymbol{u}_a|}\right|, \quad
    l_2=\left|\sqrt{\frac{\nu_l}{\zeta_l}}\log\epsilon\right|,\label{influence_region}
\end{align}
in which $\nu_l=0.1$, $\zeta_l=3.0$ are constants \textcolor{brown}{related to the flow, and $\epsilon=0.2$ is the tolerance of approximation.} The first two variables $\nu$ and $\zeta$ correspond to the diffusion and dissipation coefficient in the 1D convection-diffusion-reaction equation, from which the original derivation of Eq.~\eqref{influence_region} is made. We assume that such a simplified calculation is sufficient for our purpose and it is later proved by the parametric study of the size of the influence region in Section \ref{results}.

\begin{figure}[!htb]
\centering
\includegraphics[width=0.7\textwidth]{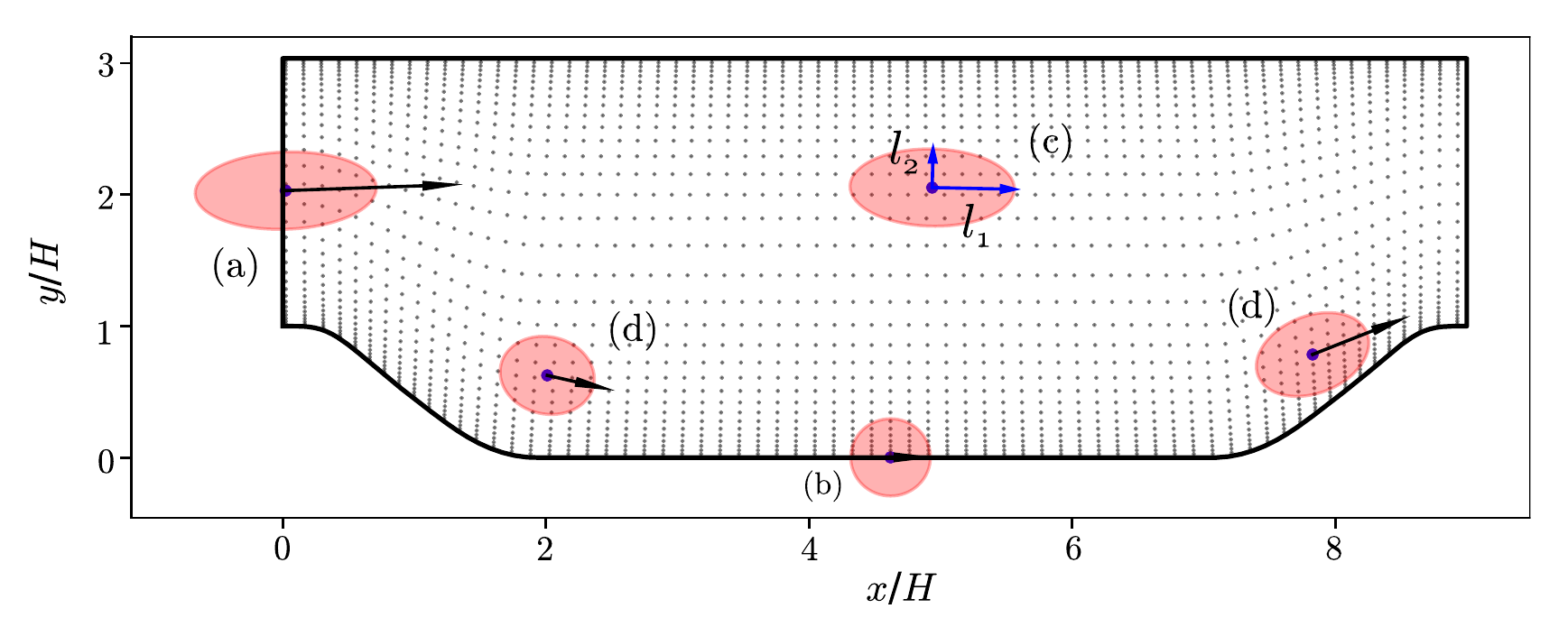}
\caption{Shape of the stencils at various locations in the flow domain. The mesh points are sampled for a clearer visualization and are presented in grey dots. The major axis $l_1$ and the minor axis $l_2$ as computed by Eq.~\eqref{influence_region} are indicated at point (c) in blue.}
\label{fig:stencils}
\end{figure}

The orientation, shape, and size of the stencil vary according to the local mean velocity based on Eq.~\eqref{influence_region}. A visualization of the stencils at various locations for the case $\alpha=1$ is presented in \autoref{fig:stencils}. Point (d) indicates how the stencils are aligned with the local mean velocity. Point (c) in the main flow has a rather large major axis due to its large local velocity. Mesh points in the vicinity of boundaries require additional treatments. For points such as (a), periodic boundary condition applies. Flow data of stencils points prior to the inlet can be accessed at the outlet region and vice versa. For points whose stencil involves the wall boundaries at the bottom and the top of the channel (point (b) in the figure), the part of the stencil outside the flow domain is ignored.

Since both the size of the stencil and the mesh density vary from location to location, the number of points in a stencil also varies. Specifically, stencils in the middle of the channel are larger while the mesh there is sparser. In contrast, stencils near walls are smaller while the mesh there is denser. By considering the stencil as a set, VCNN has the flexibility of processing stencils with a varying number of points. We took advantage of this property and adopted different strategies for training and validation. For the convenience of training, a random sampling procedure is applied to the training data set, in which 200 points are randomly drawn from the stencil at each mesh point. For the best prediction performance, we use full stencils for validation, meaning that all the data points within the stencils are used as the neural network inputs.

\subsubsection{Input features and normalization}
In this work, we select 11 features that are relevant to predicting turbulence quantities to build the feature vector for each data point. The collection of the feature vectors within the stencils is then used as the input of the neural network for predicting the turbulence quantities at the cloud center. Definitions and descriptions of them are listed in \autoref{tab:input_features}. The first two features are the relative coordinates to the cloud center. They are normalized by the distance to the cloud (stencil) center. Features 3 and 4 are the velocity relative to the cloud center, guaranteeing Galilean invariance. These four features are vectors as they will change under the rotation of the reference frame. The remaining 7 features are scalar features. The strain rate magnitude $s$ and the velocity magnitude $\textrm{u}$ provide supplementary information of the mean velocity field. The boundary cell indicator $b$ and the wall distance function $\eta$ provide related geometric information. The cell volume $\theta$ normalized by the mean is the relative cell size within the stencil. The last two features are the proximity to cloud center $r$ and the proximity in local velocity frame $r'$. The expression $\boldsymbol{u}^\top\boldsymbol{x}'$ for computing $r'$ is the inner product of the velocity and the relative coordinate to the central point. The choice of feature $r'$ is based on the property of the advection term implying that the upstream will influence the downstream. 
\begin{table}[h]
    \centering
    \begin{tabular}{lccl}
    \Xhline{4\arrayrulewidth}
         index&features& definition & description  \\
         \hline
         1&$x'$ & $\frac{x-x_0}{|\boldsymbol{x}-\boldsymbol{x}_0|+\epsilon_0}$ & relative $x$-coordinate to cloud center\\
         2&$y'$& $\frac{y-y_0}{|\boldsymbol{x}-\boldsymbol{x}_0|+\epsilon_0}$ & relative $y$-coordinate to cloud center\\
         \hline
         3&$u$ & $u_a-u_0$& relative velocity component in $x$ direction \\
         4&$v$ & $v_a-v_0$& relative velocity component in $y$ direction\\
         \hline
         5&$s$ &$||\boldsymbol{s}||$&magnitude of the strain rate tensor\\
         6&$b$ & 1(yes)/0(no) & boundary cell indicator \\
         7&$\theta$& $\theta_o/\bar{\theta}$ & cell volume \\
         8&$\textrm{u}$ &$|\boldsymbol{u}|$& relative velocity magnitude\\
         9&$\eta$ &$\min(\eta_o/l_{\delta}, 1)$&wall distance function\\
         10&$r$ &$\frac{\epsilon_r}{\sqrt{x'^2+y'^2}+\epsilon_r}$&proximity to cloud center\\
         11&$r'$ &$\epsilon_{r'}-\frac{\boldsymbol{u}^\top \boldsymbol{x}'}{|\boldsymbol{x}'|^2}$&proximity in local velocity frame\\
    \Xhline{4\arrayrulewidth}
    \end{tabular}
    \caption{Input features. $x_0$, $y_0$, $\boldsymbol{x}_0$ are the coordinates of the cloud center. $\epsilon_0(=10^{-5})$ is to avoid divided by zero. $\theta_o$ is the raw cell volume, and $\bar{\theta}$ is the mean cell volume of all the stencil points. $\eta_o$ is the raw wall distance, and $l_{\delta}$ is the prescribed boundary layer thickness scale. $\epsilon_r=0.01$ and $\epsilon_{r'}=|\boldsymbol{u}|/|\boldsymbol{x}'|$ are used to transform the function into desired range.}
    \label{tab:input_features}
\end{table}

All the features are translationally and Galilean invariant. They are invariant under a translation or a constant velocity of the reference frame because all the coordinates and velocities are relative to the cloud center. In terms of rotation, features 1-4 are not rotationally invariant. The relative coordinates and relative velocity are related to the orientation of the reference frame. Thus, the first four features require further operations to achieve rotation invariance. Features 5-11 are the scalar features, which are already rotationally invariant. For convenience of the following discussion, the feature vector composed of all 11 features is denoted by $\textbf{q}=[\textbf{x}', \textbf{u}, \textbf{c}]$, in which $\textbf{c}=[s, b, \theta, \textrm{u}, \eta, r, r']$. The input feature matrix containing feature vectors of all the sampled points in the stencil is written as
\begin{equation}
\mathcal{Q}\in \mathbb{R}^{n\times11}=
\begin{bmatrix}
\mathbf{x}_1' & \mathbf{u}_1& \mathbf{c}_1\\
\mathbf{x}_2' & \mathbf{u}_2& \mathbf{c}_2\\
\vdots & \vdots& \vdots\\
\mathbf{x}_{n}' & \mathbf{u}_{n}& \mathbf{c}_{n}
\end{bmatrix},
\end{equation}
in which $n$ is the stencil size (number of sampled points in the stencil, for training dataset $n=200$).

It is worth noting that when building the input dataset, we have already taken the numerical requirements of neural networks into consideration. For instance, $\epsilon_0$, $\epsilon_r$ and $\epsilon_{r'}$ are introduced such that the feature value is suitable for neural network training (approximately within the range $[0,1]$). However, one of the features ($s$) and both of the labels ($k$ and $\varepsilon$) are not normalized yet. In this work, we additionally cap the strain rate magnitude to 3 for simplicity. The turbulent kinetic energy is normalized by the reference velocity $k\propto\frac{k_\text{o}}{(U_{\textrm{ref}})^2}$ ($k_\text{o}$ represents the original turbulent kinetic energy in the flow simulation and $k$ is the normalized value to be predicted by neural networks) and scaled by turbulence intensity $I_t=0.2$ such that the normalized value falls roughly between 0 and 1. The normalized $k$ follows
\begin{equation}
    k=\frac{k_\text{o}}{(U_{\textrm{ref}}I_t)^2}.
\end{equation}
The normalization of $k$ involves the reference velocity $U_{\textrm{ref}}$. In our case it can be regarded as the relative velocity with regards to the channel and thus it is Galilean invariant.

The turbulent kinetic energy dissipation rate is transformed by taking the logarithm $\varepsilon\propto\log_{10}{\varepsilon_\text{o}}$ considering its distribution in a broad range and its positivity:
\begin{equation}
    \varepsilon=\log_{10}{\varepsilon_\text{o}},
\end{equation}
in which $\varepsilon_\text{o}$ represents the original turbulence kinetic energy dissipation rate and $\varepsilon$ is the normalized dissipation rate to be predicted by neural networks. It is further scaled by a constant 0.5 to attain the proper range that is suitable for training.

\subsection{Neural network architecture and training}
\begin{figure}[!htb]
\centering
\includegraphics[width=0.9\textwidth]{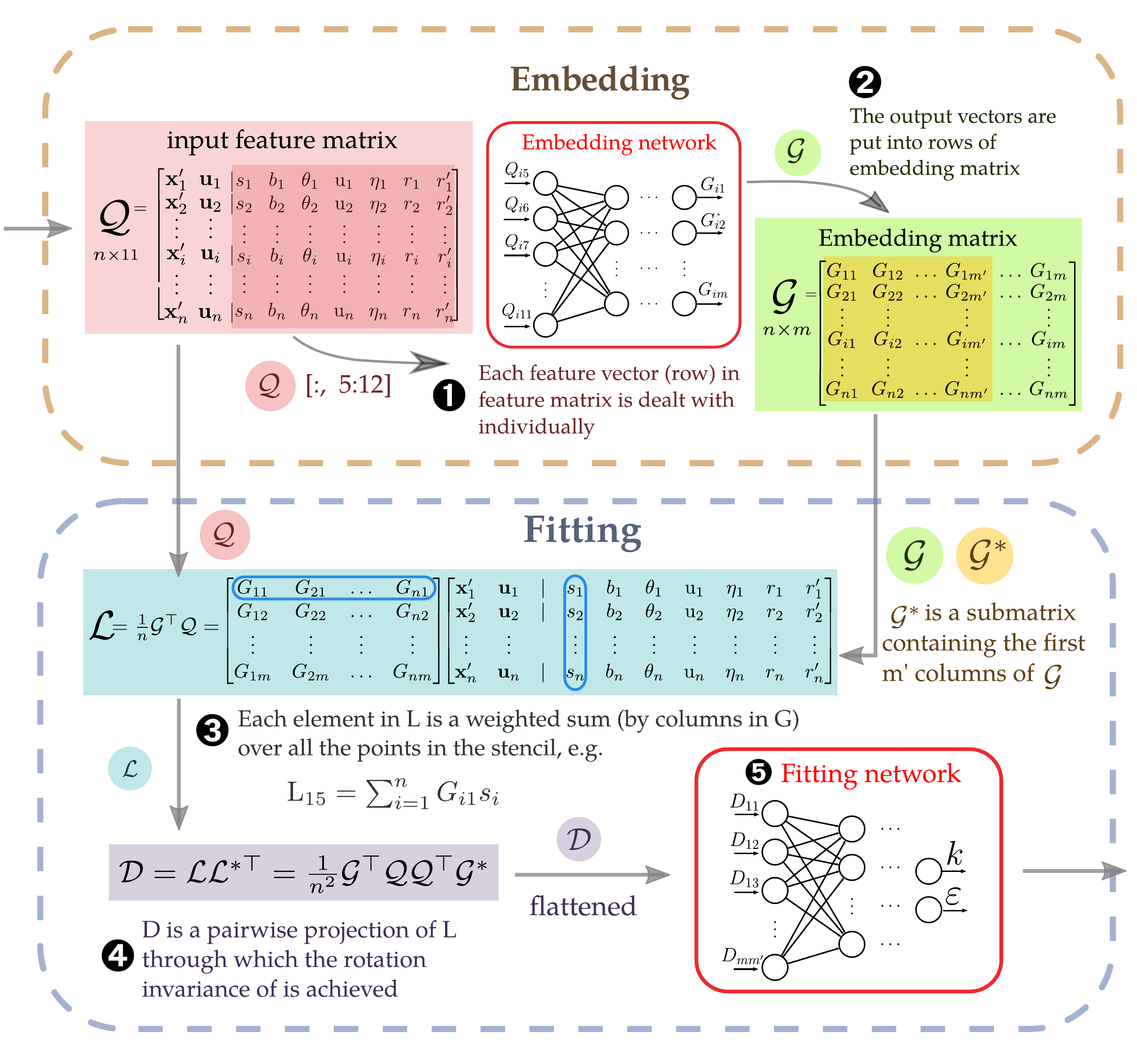}
\caption{Schematics for the vector-cloud neural network. The neural network takes the feature vectors (rows in $\mathcal{Q}$) as the inputs and provides predictions on turbulence quantities $k$ and $\varepsilon$. In the first embedding stage, the scalar features from each row of the input feature matrix are fed into the embedding network, in which they are transformed by the embedding network into rotation invariant embedding matrix $\mathcal{G}$ and its submatrix $\mathcal{G^*}$. In this stage, all the elements in the stencil are treated identically and individually. In the second fitting stage, information from each stencil point is integrated permutation invariantly into nonlocal features in $\mathcal{L}$ and $\mathcal{L}^*$ through a weighted sum. A pairwise projection is performed to achieve rotation invariance of the vector features in $\mathcal{Q}$. $\mathcal{D}$ is flattened into a vector and fed into the fitting network. The outputs of the fitting network are the prediction on turbulence quantities $k$ and $\varepsilon$.}
\label{fig:network_schematics}
\end{figure}

The VCNN architecture is composed of two multilayer perceptron (MLP) sub-networks. An overall schematic of the network architecture is provided in \autoref{fig:network_schematics} and major steps are numbered. The first sub-network is referred to as the embedding network, which aims at extracting higher-level features that are correlated to the turbulence quantities (step \textcircled{1}).
The embedding network has a $m$-dimensional output and it represents $m$ one-dimensional embedding functions $\phi_1, \dots, \phi_m$. These embedding functions are evaluated at each cloud center $i$ using the frame-independent scalar features $\textbf{c}_i, i=1, 2, \dots, n$ as inputs and provide $G_{ij}=\phi_j(\textbf{c}_i), j=1, 2, \dots, m$. 
Through training, the embedding functions will embody important nonlinear relationships between features in vector $\textbf{c}_i$. The embedding functions evaluated at all the samples in the same stencil are arranged into a matrix with its $n$ rows corresponding to $n$ points in the stencil and $m$ columns corresponding to $m$ embedding functions (step \textcircled{2}); see $\mathcal{G}=(G_{ij})\in \mathbb{R}^{n \times m}$ in Eq.~\eqref{projection}, appearing in the transposed form.

It is crucial that samples in the stencil are processed independently and identically in the embedding network. The $m$ embedding functions evaluated at each stencil point are only dependent on the features of the point itself and they are irrelevant to other stencil points. This is to guarantee that the final prediction is permutation invariant with the index of samples.
The embedded feature matrix $\mathcal{G}$ are frame-independent since they are derived from frame-independent features $\textbf{c}$.

As a nonlocal model, the information contained in the single stencil points needs to be integrated into features of the entire stencil. This is achieved through averaging the input features $\textbf{q}_i$ according to their embedded weighs $G_{ij}$ (step \textcircled{3}):
\begin{equation}\label{projection}
    \mathcal{L}=\frac{1}{n}\mathcal{G}^\top\mathcal{Q}=\begin{bmatrix}
G_{11}& G_{21}& \dots &G_{n1}\\
G_{12}& G_{22}& \dots &G_{n2}\\
\vdots & \vdots& \ddots& \vdots\\
G_{1m}& G_{2m}& \dots &G_{nm}
\end{bmatrix}\begin{bmatrix}
\mathbf{q}_1^\top\\
\mathbf{q}_2^\top\\
\vdots\\
\mathbf{q}_n^\top\\
\end{bmatrix}=\frac{1}{n}\begin{bmatrix}
\sum_{i=1}^nG_{i1}\textbf{q}_i^\top\\
\sum_{i=1}^nG_{i2}\textbf{q}_i^\top\\
\vdots\\
\sum_{i=1}^nG_{im}\textbf{q}_i^\top
\end{bmatrix}\in\mathbb{R}^{m \times 11}.
\end{equation}
Eq.~\eqref{projection} can be regarded as $m$ forms of weighted summation of the input features. The operation is permutation invariant since the weights are dependent of single stencil points $G_{ij}=\phi_j(\textbf{c}_i)$ as emphasized earlier. It follows directly: 
\begin{align}
    \sum_{i=1}^nG_{ij}\textbf{q}_i^\top&=\sum_{i=1}^n\phi_j(\textbf{c}_i)\textbf{q}_i^\top \quad \text{for} \quad j=1,\dots, m.
\end{align}
Each of the elements in $\mathcal{L}$ is a summation over functions of each sample in the stencil. Switching orders of any two of the samples will not influence the results.

Upon introducing $\mathbf{q}_i$ (thus $\textbf{x}'$ and $\textbf{u}$) into the calculation of $\mathcal{L}$, rotation invariance is violated. This can be solved by applying a pairwise projection (step \textcircled{4}):
\begin{equation}
    \mathcal{D}=\mathcal{L}\mathcal{L}^\top=\frac{1}{n^2}\mathcal{G}^\top\mathcal{Q}\mathcal{Q}^\top\mathcal{G}\in\mathbb{R}^{m \times m}.
\end{equation}
The pairwise projection $\mathcal{Q}\mathcal{Q}^\top$ means projecting each feature vector to every other feature vector. In this way, the resultant matrix captures the correlation among samples that is independent of any rotation of the reference frame. As both the averaging based on the embedded weights and the pairwise projection are linear transformations, the associative law of linear transformation applies. The matrix $\mathcal{D}$ is both permutation invariant and rotation invariant. As the input of the fitting network, $\mathcal{D}$ also guarantees the desired invariant properties of the network. In the fitting network, the model's final outputs, the turbulence quantities $k$ and $\varepsilon$ are predicted (step \textcircled{5}). 

\begin{figure}[!htb]
\centering
\includegraphics[width=1\textwidth]{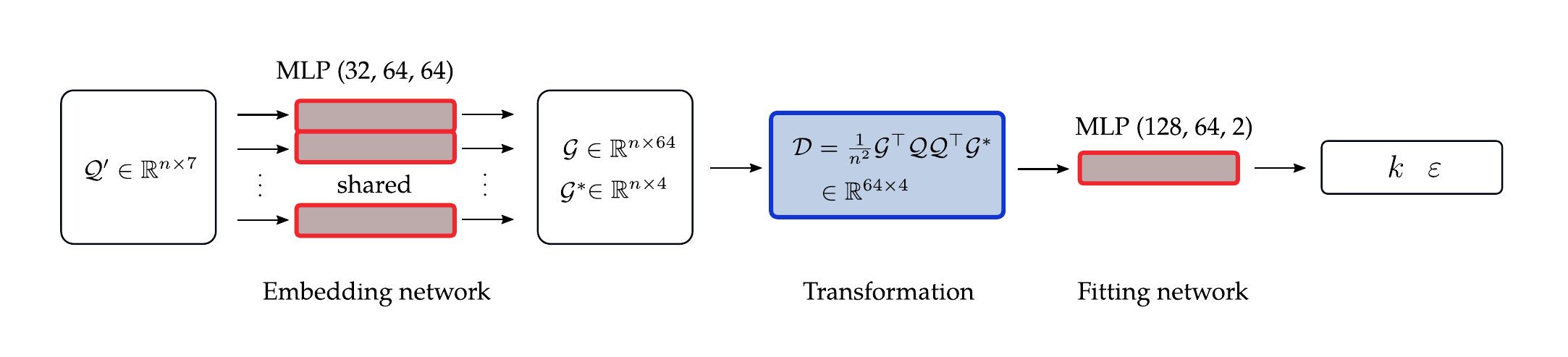}
\caption{Hyperparameters of the VCNN}
\label{fig:hyperparameters}
\end{figure}

\textcolor{brown}{
In our implementation, a submatrix $\mathcal{G}^*\in\mathbb{R}^{n \times m'}$ of the full embedding matrix $\mathcal{G}\in\mathbb{R}^{n \times m}$ is used in the linear transformation in order to reduce the dimension of $\mathcal{D}$ and save computational cost while all the invariance properties are still preserved:
\begin{equation}
    \mathcal{D}=\mathcal{L}\mathcal{L}^{*\top}=\frac{1}{n^2}\mathcal{G}^\top\mathcal{Q}\mathcal{Q}^\top\mathcal{G}^*\in\mathbb{R}^{m \times m'}.
\end{equation}
We choose $m'=4$ so that $\mathcal{G}^*$ contains the first 4 columns of $\mathcal{G}$. The exact hyperparameters of the network used in this work are shown in \autoref{fig:hyperparameters}.}

The loss function used for training the neural network naturally contains the error of outputs $k$ and $\varepsilon$. Besides, the ultimate target turbulence viscosity $\nu_T$ in Eq.~\eqref{nut_ke} is also included as a supplement term for guiding the network towards better prediction performance. The complete loss function then follows
\begin{equation}
    R(\theta)=\sum_{i=1}^N [(\hat{k}_i-k_{i_n})^2+ (\hat{\varepsilon}_i-\varepsilon_{i_n})^2+\lambda (\hat{\nu}_{T_i}-\nu_{T_i})^2],
\end{equation}
where $\lambda$ is a weighting factor balancing the two normalized error terms of $k$ and $\varepsilon$ with the error term of $\nu_T$; $\hat{k}$, $\hat{\varepsilon}$ and $\hat{\nu_T}$ represent predictions made by the neural network model. Adam optimization algorithm~\cite{kingma2014adam} is used for training neural network parameters with an initial learning rate $10^{-3}$. The learning rate decreases by a factor of 0.7 for every 600 steps. The network is trained until the training error reaches the lowest and the validation error stays at a low level. The number of epochs is around 2000. The construction of the neural network and its training are implemented using the PyTorch machine learning library~\cite{NEURIPS2019_9015}. 

\subsection{Coupling with RANS equation solver}
The neural network described in the previous section should work as a usual turbulence model in flow simulation. In this work, the neural network-based turbulence model is coupled with the SimpleFOAM solver from the OpenFOAM package. The network requests mean flow solutions (mean velocities $\boldsymbol{u}$ and mean strain rate $s$), geometry data (relative coordinates $\boldsymbol{x}'$, cell volume $\theta$, boundary and wall distance $b$ and $\eta$, proximity to cloud center $r$ and $r'$) from the OpenFOAM solver and returns turbulence quantities $k$, $\varepsilon$. 

\begin{figure}[!htb]
\centering
\includegraphics[width=1\textwidth]{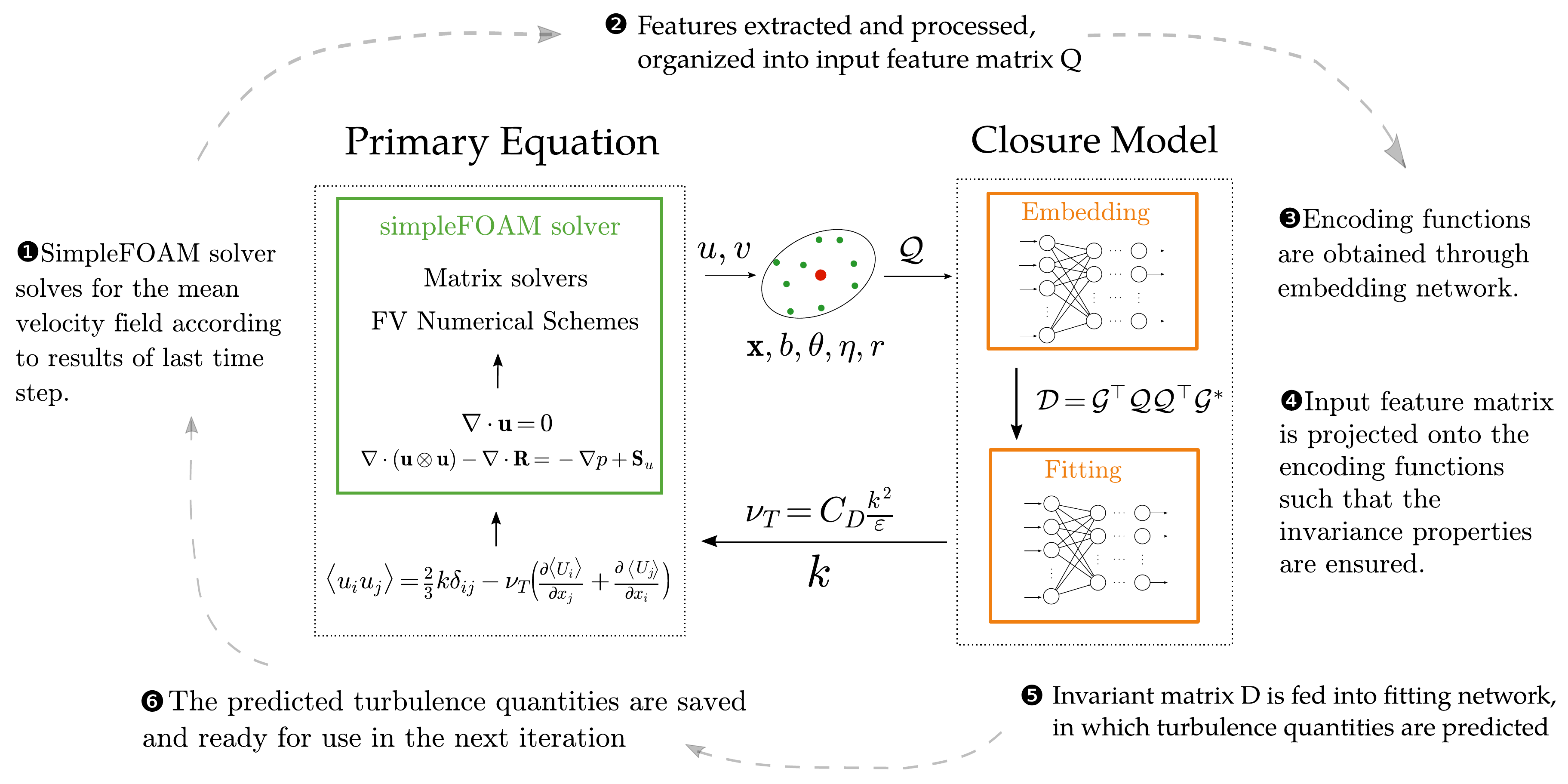}
\caption{Schematics of the coupled solver. Steps in each iteration are: (1) SimpleFOAM solver solves for the primary equation. (2) Compute required features. (3)-(6) Neural network makes a prediction on turbulence quantities}
\label{fig:coupled_schematics}
\end{figure}

The pipeline of data exchange is presented in \autoref{fig:coupled_schematics} and the steps for a coupled solver are as follows:
\begin{enumerate}[(1)]
    \item The mean flow field and the turbulence quantities are initialized by the initial condition. The absolute wall distance $\eta_o$ of all the mesh points is computed.
    \item SimpleFOAM solver reads the initial data or results from the last time step and solves for the mean flow quantities, e.g., flow field, pressure. Data required by the neural network such as the strain rate magnitude, cell volume, etc., are also calculated.
    \item The data is read and processed as described in Section~\ref{data_processing}, including collecting the stencil point, calculating features and normalization. The processed data is put into the input matrix $\mathcal{Q}$ and fed into the trained neural network.
    \item The data is processed first through the embedding network, then transformed to achieve the permutation invariance and frame-independence. Finally, the prediction on turbulence quantities is given by the fitting network and is written into OpenFOAM format.
    \item Steps 2--4 are repeated until the simulation reaches convergence.
\end{enumerate}

\section{Results}\label{results}

\subsection{Performance in uncoupled and coupled setup}
We test the performance of the trained neural network model in both uncoupled and coupled setups.
Specifically, in the uncoupled setup, the neural network model uses the converged mean flow field obtained from traditional flow simulation to predict the turbulence quantities, which is similar to how the network is trained. In a coupled setup, the neural network model is coupled with a traditional RANS equation solver. The coupled solver starts iteration from a given initial condition, that is, an unconverged flow field. During the simulation, the neural network model makes predictions on turbulence quantities in each iteration. The simulation runs until it reaches convergence. In both cases, the neural network prediction is compared with that of the standard $k$--$\varepsilon$ turbulence model. It was found that, in the uncoupled setup, the neural network model performed very well in interpolation flow cases, but the prediction worsened drastically in extrapolation cases. In the coupled setup, the network model performance was comparably good as it was in the uncoupled setup.

\subsubsection{Uncoupled results}
\begin{figure}[!htb]
\centering
{\includegraphics[height=0.05\textwidth]{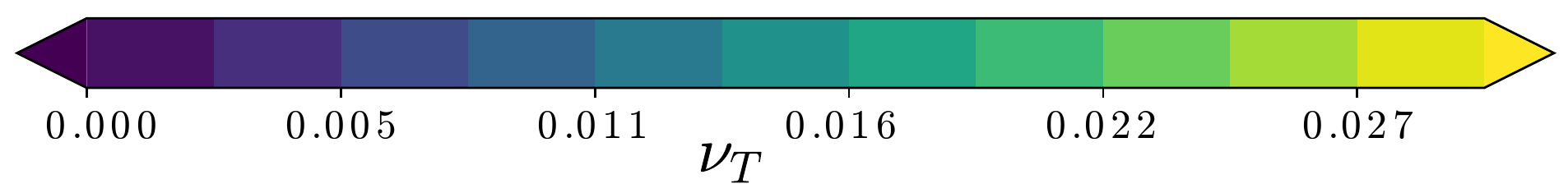}\vspace{0pt}}
\subfloat[$k$--$\varepsilon$ model, $\alpha = 1.45$]
{\label{fig:1.45_contour_truth}\includegraphics[height=0.15\textwidth]{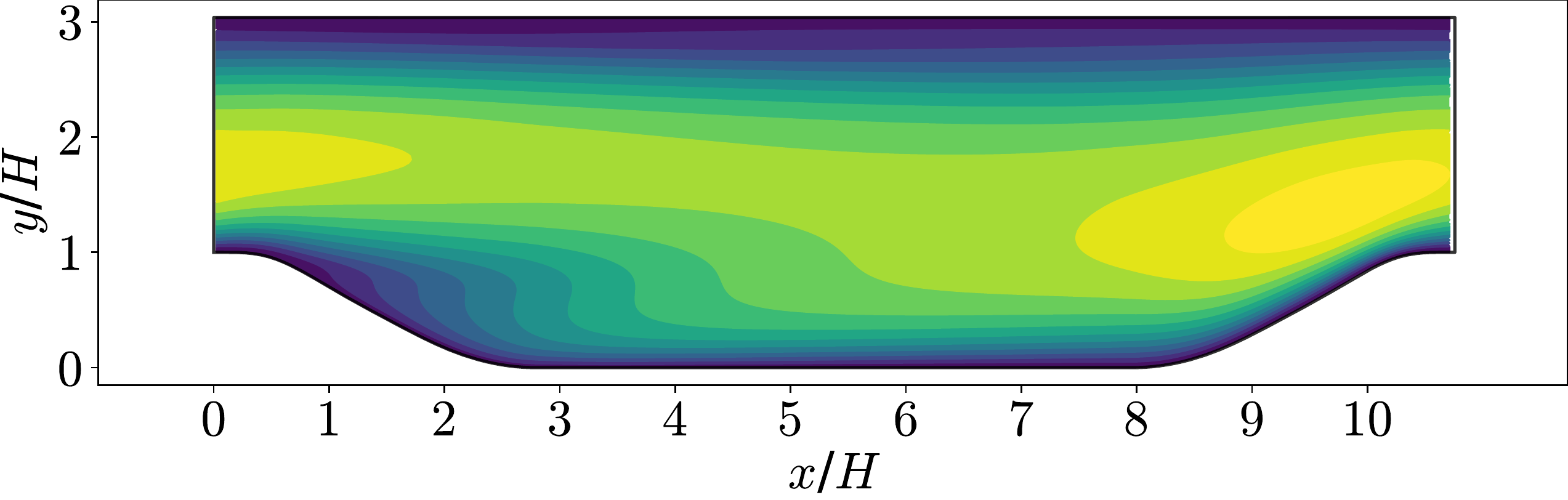}\hspace{3pt}}
\subfloat[neural network prediction, $\alpha = 1.45$]
{\label{fig:1.45_contour_pre}\includegraphics[height=0.15\textwidth]{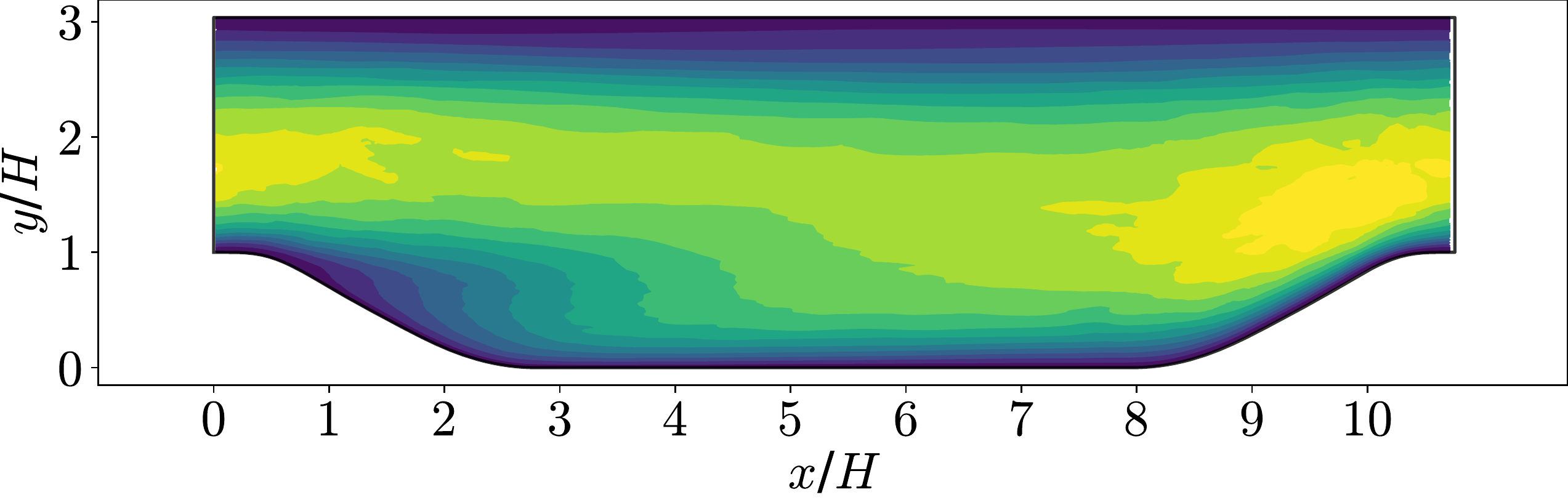}}\\
{\vspace{5pt}\includegraphics[height=0.025\textwidth]{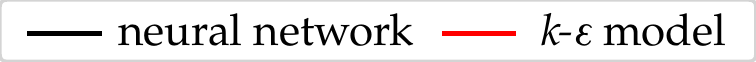}\vspace{5pt}}\\
\subfloat[neural network prediction, $\alpha=1.45$]
{\label{fig:1.45_profile}\includegraphics[height=0.15\textwidth]{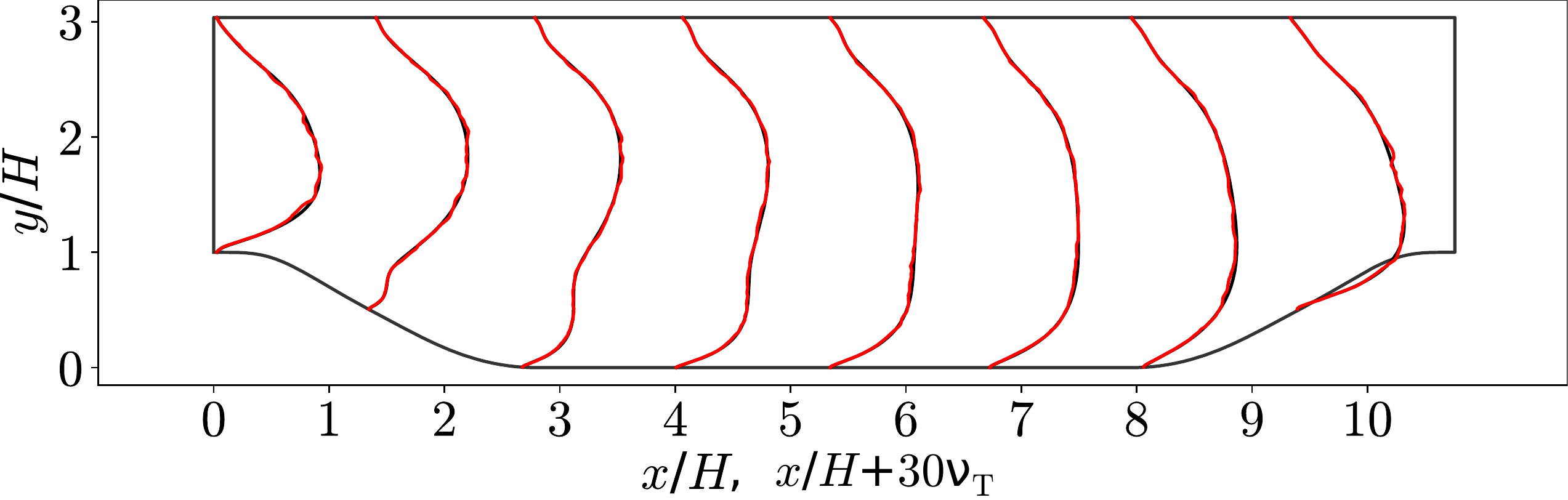}}
\caption{Comparison of turbulence viscosity $\nu_T$ on flow case $\alpha=1.45$. Panel (a) and (b) show the contour plot of $\nu_T$ provided by the $k$--$\varepsilon$ model and neural network, respectively. Panel (c) compares $\nu_T$ profiles along 8 cross-sections.}
\label{fig:interpolation-compare}
\end{figure}

In the interpolation testing case $\alpha=1.45$, the prediction of the neural network is highly consistent with the results of $k$--$\varepsilon$ model. Although the flow case $\alpha=1.45$ is not included in the training set, the contour of $\nu_T$ with $k$--$\varepsilon$ (\cref{fig:1.45_contour_truth}) and that with neural network model (\cref{fig:1.45_contour_pre}) are nearly identical in terms of their pattern and magnitude. When comparing their cross-section profiles (\cref{fig:1.45_profile}) along the channel, it is also clear that the two profiles overlapped each other. The neural network prediction is only inferior to that of $k$--$\varepsilon$ model in terms of its smoothness.

\begin{figure}[!htb]
{\hspace{3pt}\includegraphics[height=0.05\textwidth]{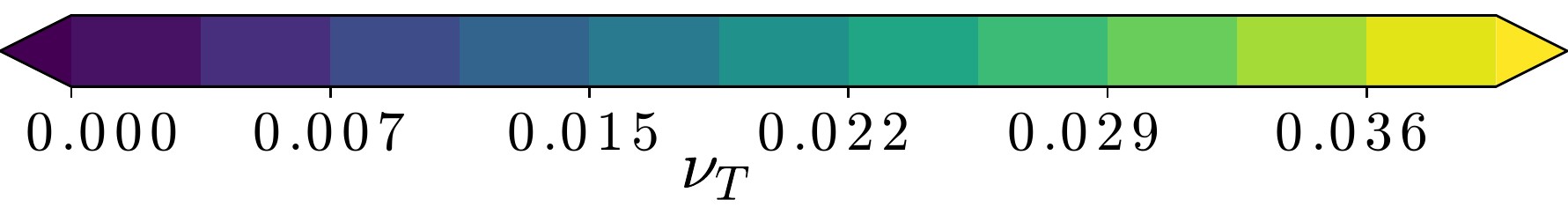}\hspace{20pt}\vspace{5pt}}
{\includegraphics[height=0.05\textwidth]{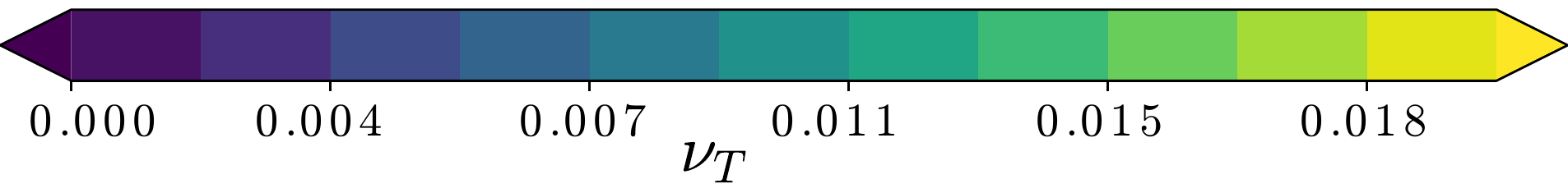}\hspace{15pt}\vspace{5pt}}\\
\centering
\subfloat[$k$--$\varepsilon$ model, $\alpha = 0.9$]
{\includegraphics[height=0.15\textwidth]{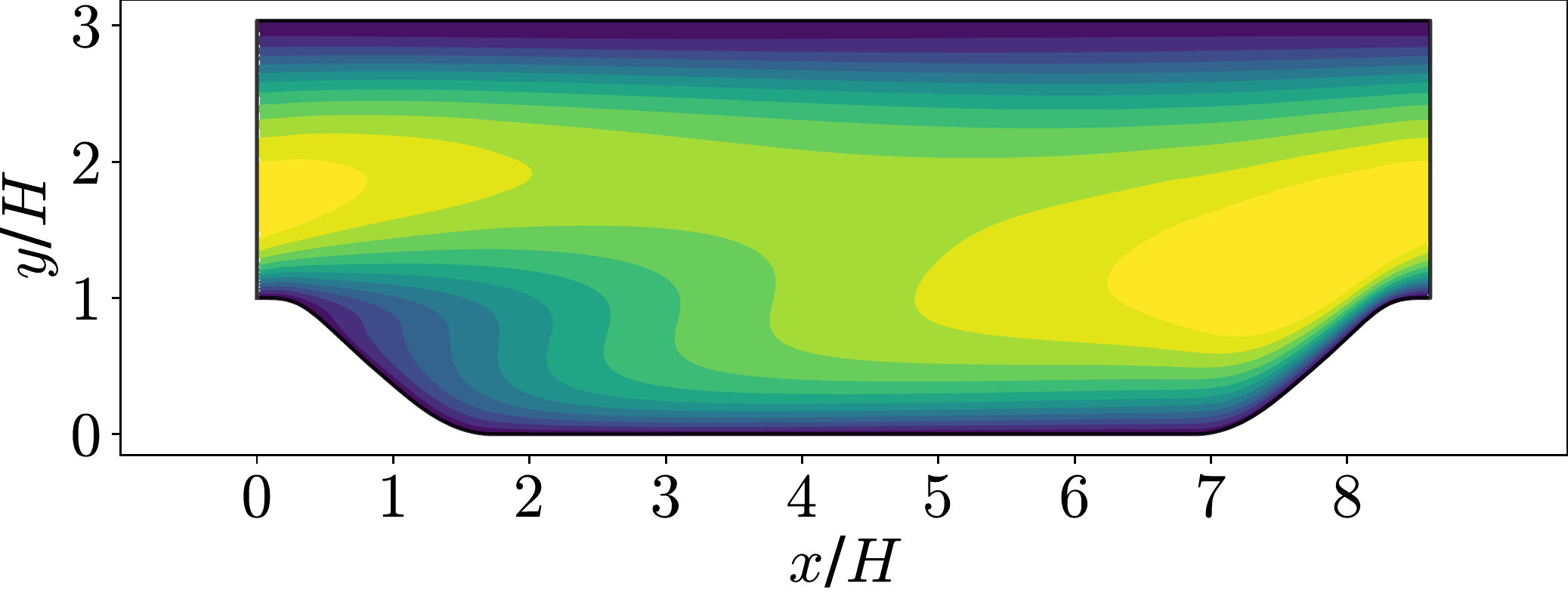}\hspace{3pt}}
\subfloat[$k$--$\varepsilon$ model, $\alpha = 2.1$]
{\includegraphics[height=0.15\textwidth]{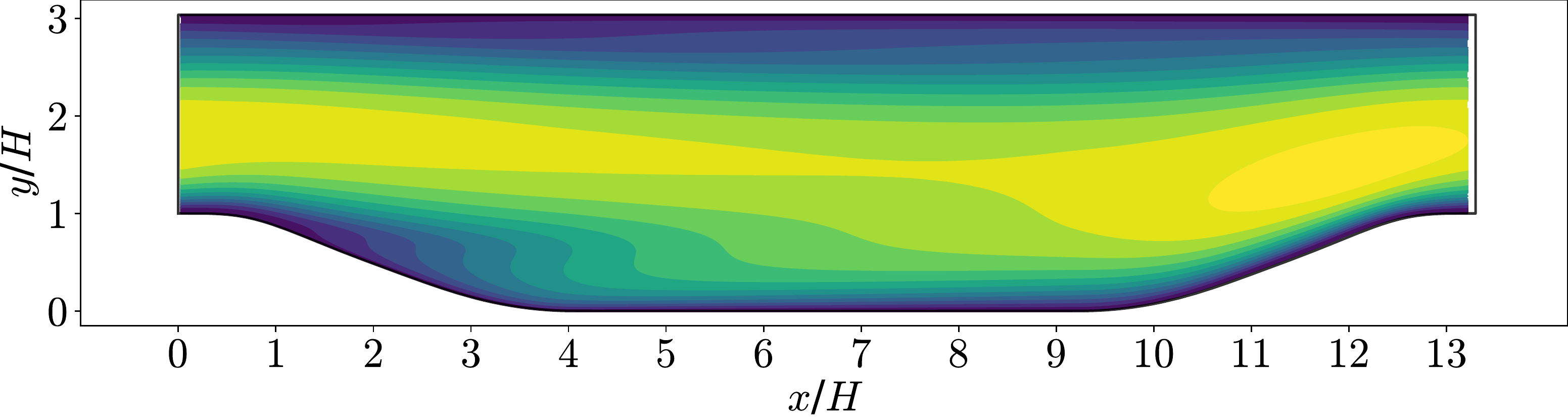}}\\

\subfloat[neural network prediction, $\alpha = 0.9$]
{\includegraphics[height=0.15\textwidth]{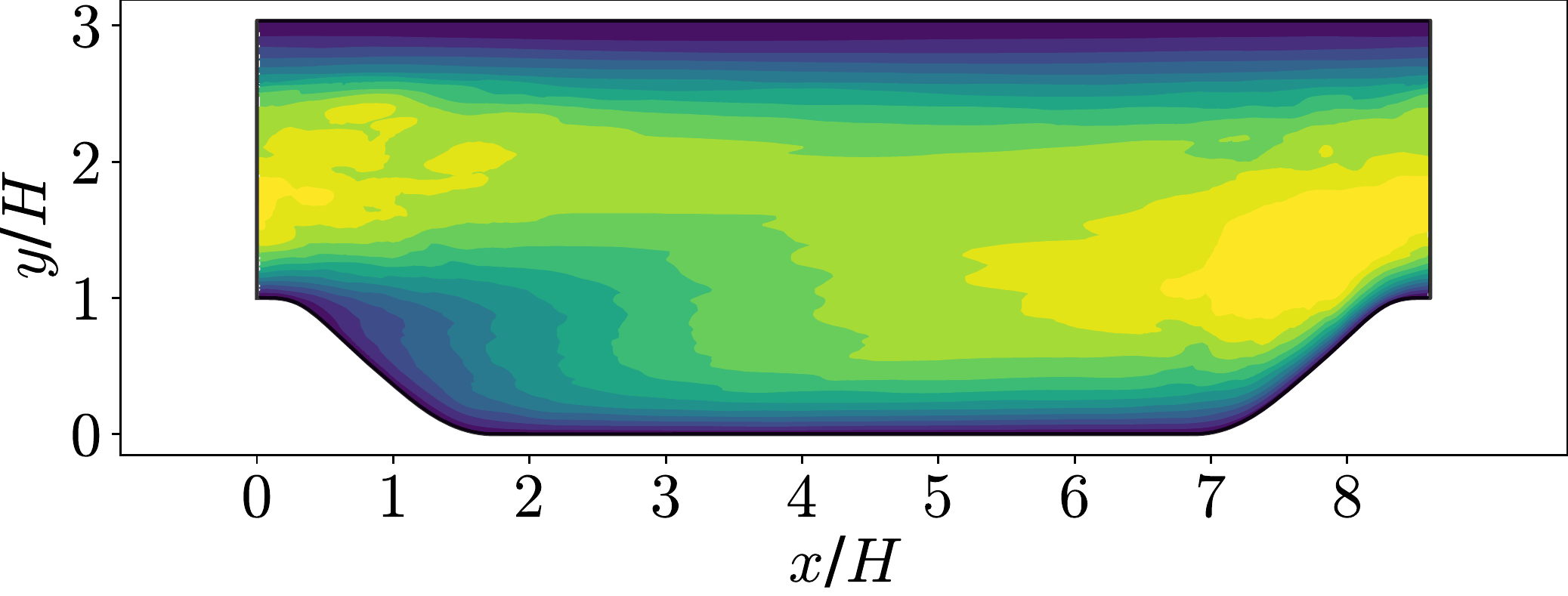}\hspace{3pt}}
\subfloat[neural network prediction, $\alpha = 2.1$]
{\includegraphics[height=0.15\textwidth]{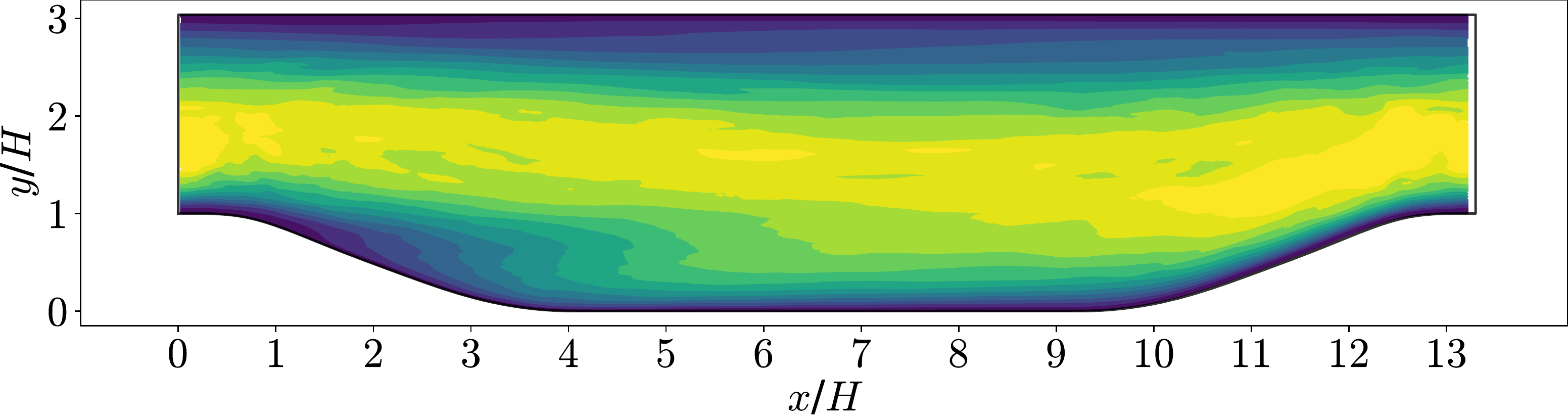}}\\
{\vspace{5pt}\includegraphics[height=0.025\textwidth]{profile_a=2.1_legend.pdf}\vspace{5pt}}\\

\subfloat[cross-section, $\alpha = 0.9$]
{\includegraphics[height=0.15\textwidth]{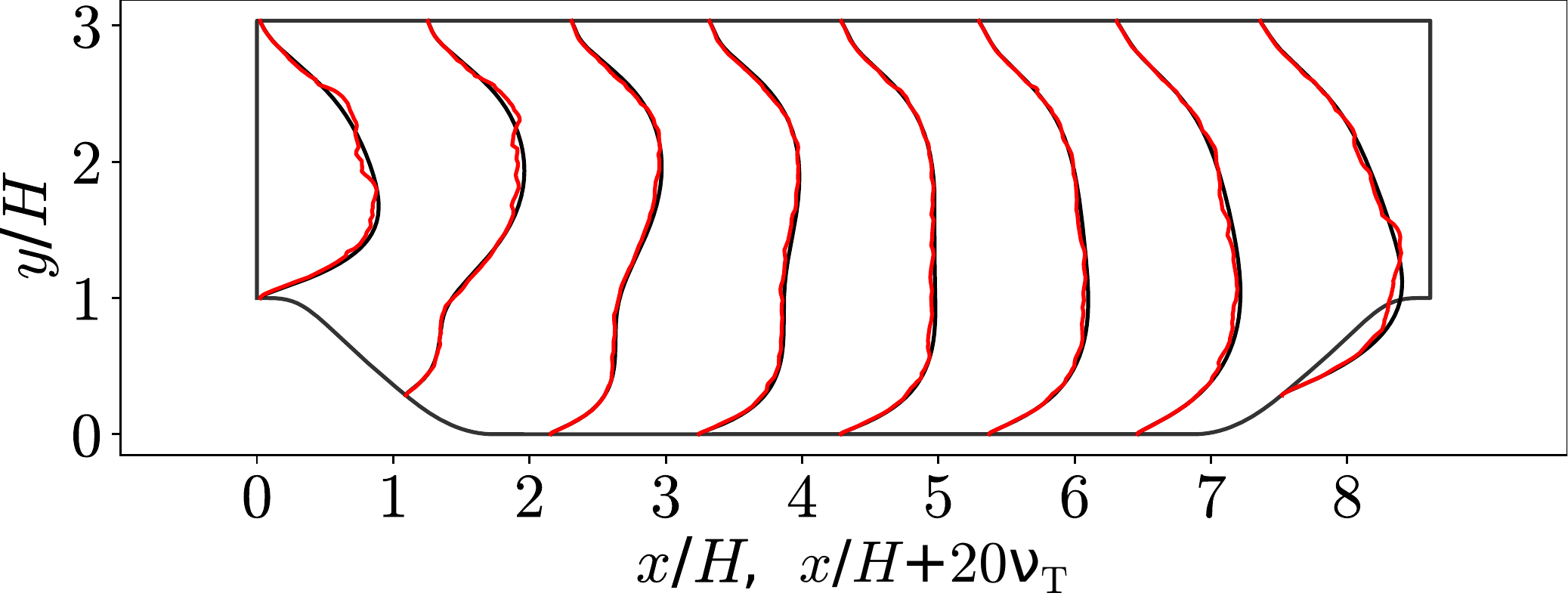}
\hspace{2pt}}
\subfloat[cross-section, $\alpha = 2.1$]
{\includegraphics[height=0.15\textwidth]{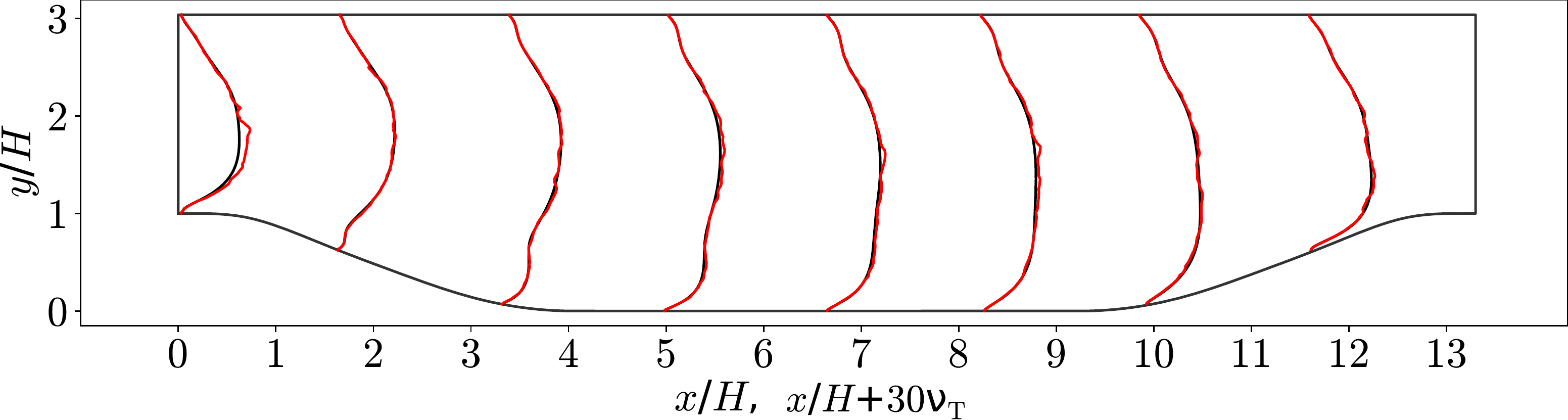}}\\
\caption{Comparison of the contour and the cross-section profiles of turbulence viscosity $\nu_T$ in the extrapolation cases $\alpha=0.9$ and $\alpha=2.1$.
}
\label{fig:extrapolation-compare}
\end{figure}

The neural network model performed less satisfactorily in extrapolation cases. The performance deteriorated rapidly in the testing cases with slopes further from the training set. Simulation results in \autoref{fig:extrapolation-compare} show the prediction capacity of the neural network model on mild extrapolation cases. In the flow cases $\alpha=0.9$ and $\alpha=2.1$, the network prediction is close to that of the reference results provided by the $k$--$\varepsilon$ model. The overall magnitude of $\nu_T$ in the neural network prediction agrees well with the $k$--$\varepsilon$ model, although there are more evident oscillations observed near the inlet and outlet of the channel, and there is also disagreement of pattern in the middle section in flow case $\alpha=2.1$.

For a more quantitative analysis of the model performance, an overall metric of validation (prediction) error rate is introduced:
\begin{equation}\label{eq:error}
    \text{error rate}=\frac{\sqrt{\sum_{i=1}^N|\nu_{T_i}-\hat{\nu}_{T_i}|^2}}{\sqrt{\sum_{i=1}^N|{\nu}_{T_i}|^2}},
\end{equation}
where $\hat{\nu_T}$ is the neural network prediction, and ${\nu}_T$ refers to the results of $k$--$\varepsilon$ model (regarded as the ground truth). The errors of all mesh points are summed and the total error is normalized by the sum of squared ground truth results.

\begin{figure}[!htb]
\centering
\includegraphics[width=0.7\textwidth]{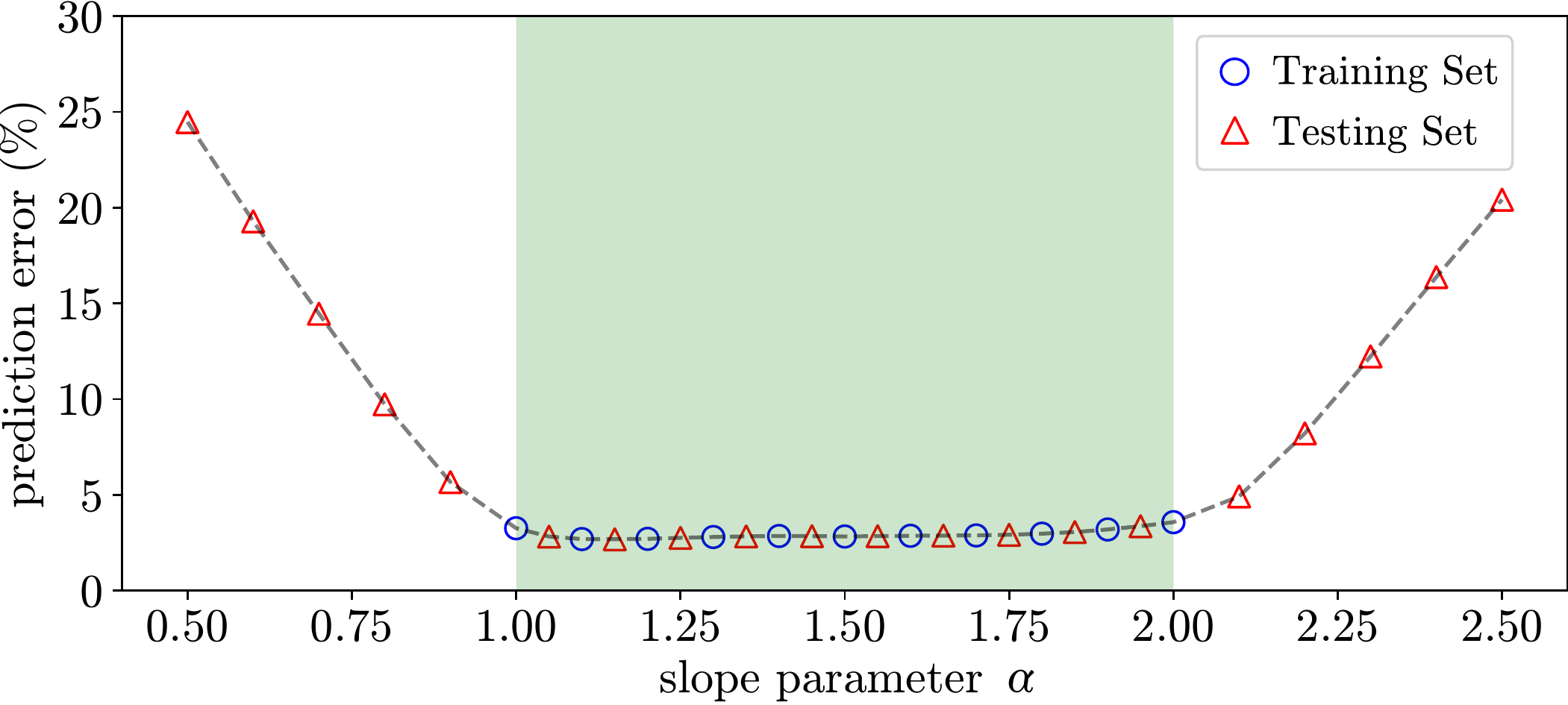}
\caption{Prediction error defined in Eq.~\eqref{eq:error} over different slope parameters. Shaded (green) background indicates the range of slope parameter covered by the training set (marked with blue circles). Red triangles represent the testing flow cases, including both interpolation and extrapolation cases.}
\label{fig:validation_error}
\end{figure}

The neural network model shows good performance in the interpolation regime but in the extrapolation regime, the performance is less satisfactory. This finding is clearly shown in \autoref{fig:validation_error}, where the prediction error rate is plotted against the slope parameters. Within the training set range, the prediction error of the neural network model is as low as the training error (around 3\%). The prediction error rate grows rapidly when the slope parameter is reduced from $1.0$ to $0.5$. Similarly, the error rate increases when the slope parameter goes beyond $2.0$, although the error rate of the case $\alpha=2.5$ is slightly lower than that of the case~$\alpha=0.5$.

\subsubsection{Coupled results}
\begin{figure}[!htb]
\centering
{\vspace{5pt}\includegraphics[height=0.025\textwidth]{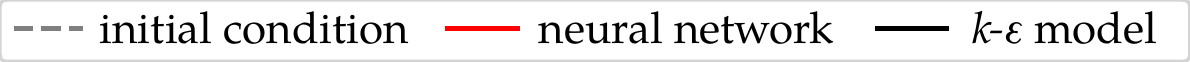}\vspace{5pt}}\\
\subfloat[$\nu_T$ profiles, $\alpha = 0.9$]
{\includegraphics[height=0.15\textwidth]{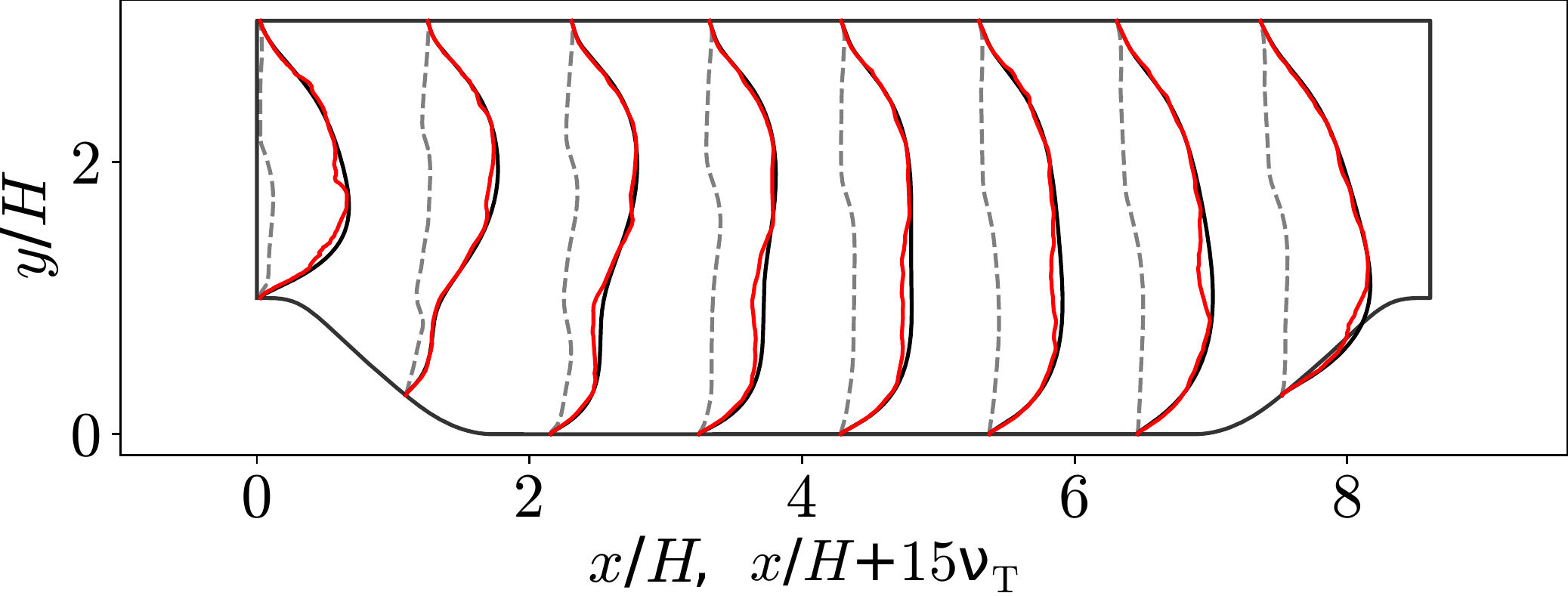}\hspace{5pt}}
\subfloat[$\nu_T$ profiles, $\alpha = 1.45$]
{\includegraphics[height=0.15\textwidth]{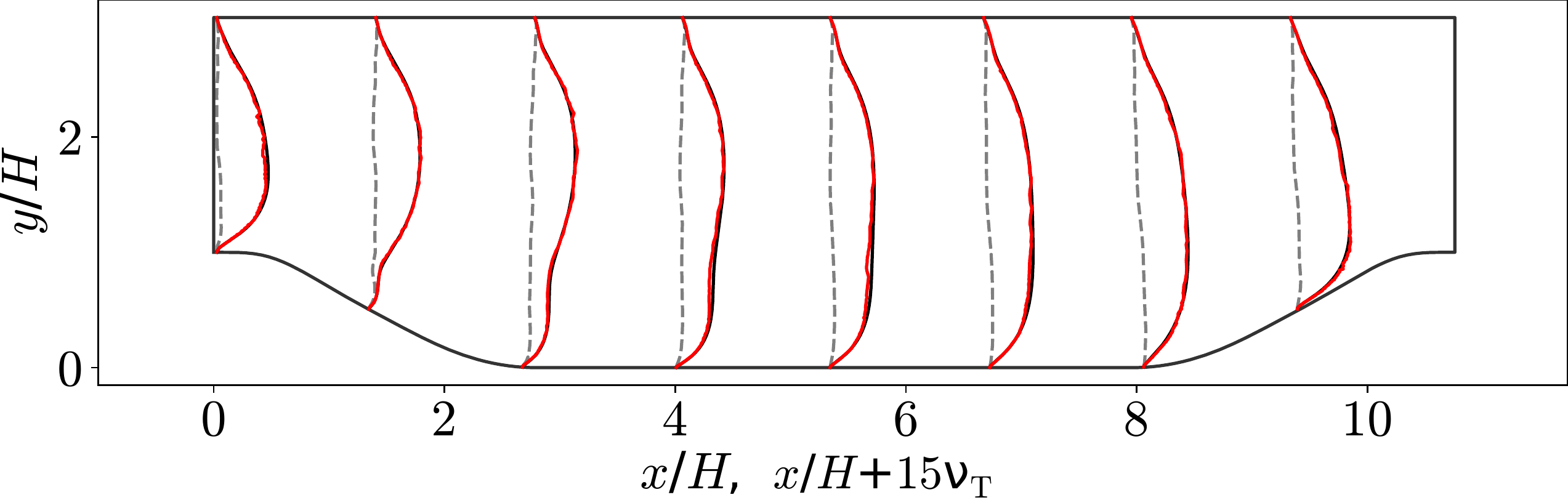}\hspace{0pt}}\\
\subfloat[$u$ profiles, $\alpha = 0.9$]
{\includegraphics[height=0.15\textwidth]{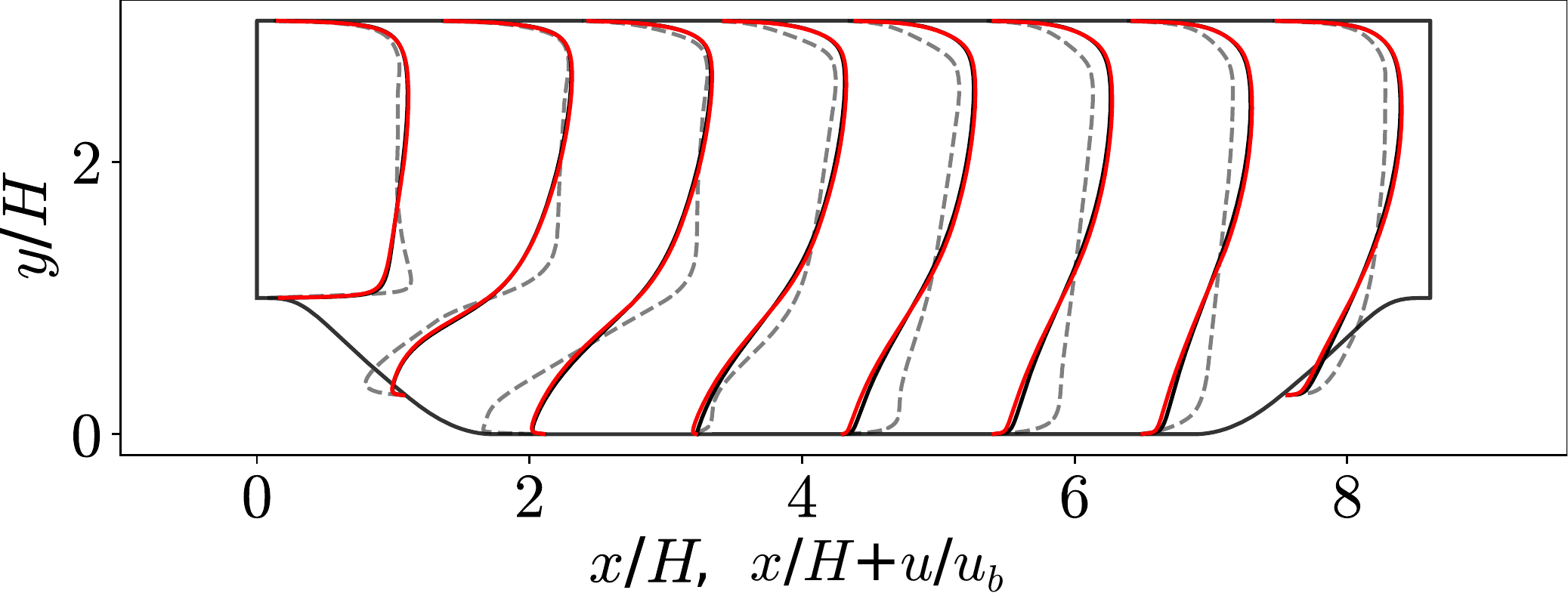}\hspace{5pt}}
\subfloat[$u$ profiles, $\alpha = 1.45$]
{\includegraphics[height=0.15\textwidth]{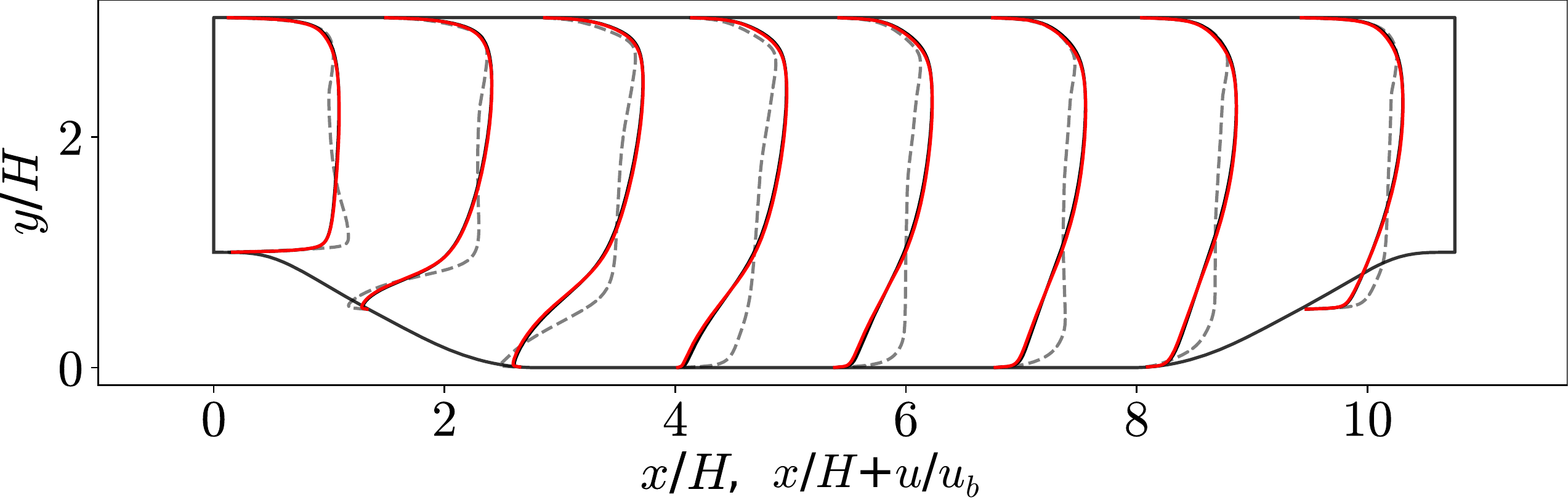}\hspace{0pt}}\\
\subfloat[$v$ profiles, $\alpha = 0.9$]
{\includegraphics[height=0.15\textwidth]{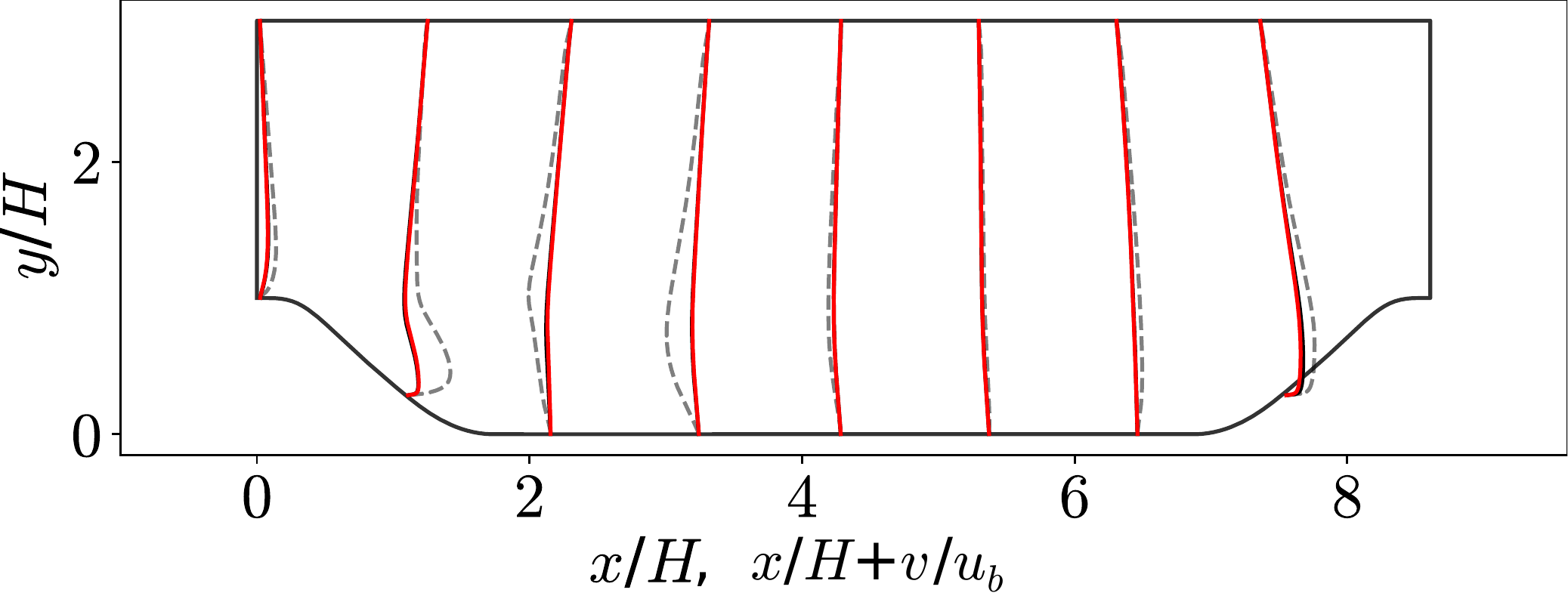}\hspace{5pt}}
\subfloat[$v$ profiles, $\alpha = 1.45$]
{\includegraphics[height=0.15\textwidth]{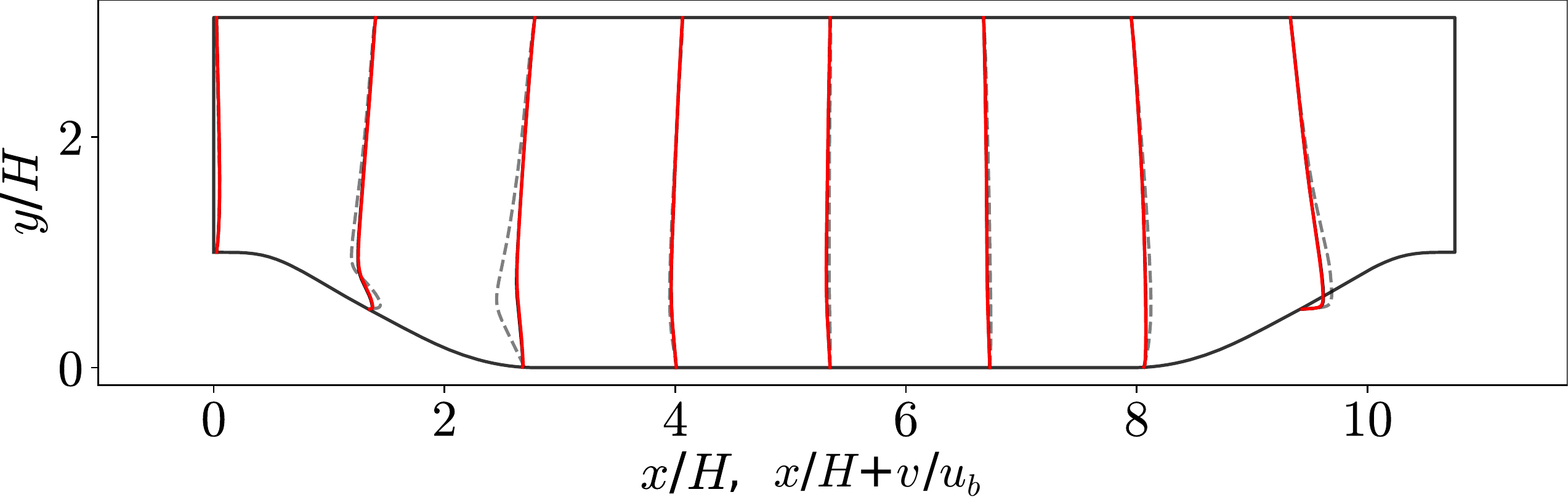}}\\
\caption{Comparison of the velocity and eddy viscosity profiles. The initial condition marked in the grey dashed line is the corresponding baseline $k$--$\varepsilon$ model simulation at $500^\textrm{th}$ iteration (with the first iteration started from uniform flow fields).
}
\label{fig:coupled_comparison}
\end{figure}

In the coupled setup, the neural network model performs comparably well as in the uncoupled setup in terms of the model accuracy. The converged fields in one interpolation and one mild extrapolation flow case of the coupled solver with the neural network-based turbulence model and the $k$--$\varepsilon$ model are presented in \autoref{fig:coupled_comparison}. The cross-section profiles of the eddy viscosity $\nu_T$ and the mean velocity in $x$ and $y$ direction ($u$ and $v$) are shown. We note that the difference between the neural network-based turbulence model and the baseline $k$--$\varepsilon$ model is very small. Especially in terms of the mean flow field (the final results of RANS simulation), the difference is barely discernible.

More importantly, the network has good stability and the coupled solver is capable of reaching a steady state when initialized with an unconverged flow field at the early stage. The initial condition of the simulation is shown in grey dashed lines in \autoref{fig:coupled_comparison}, and corresponds to the baseline simulation at the $500^\textrm{th}$ iteration. There is a large discrepancy between the initial condition and the final converged results, especially in terms of the eddy viscosity. Such an unconverged flow field is not included in any training cases. Even in an unseen flow field, the neural network-based turbulence model is capable of providing reasonable prediction without causing divergence of the primary RANS equation and brings the simulation to a convergence that is very close to the $k$--$\varepsilon$ model.

The starting point of $500^\textrm{th}$ iteration is found through grid search, i.e., we tested the neural network-coupled RANS solver starting from the converged flow field all the way down to the $1^\textrm{st}$ iteration. Tests starting from iterations earlier than the $500^\textrm{th}$ hardly succeeded. This is because the difference between the mean flow fields in the initial stage and those in the final stage is too large, and consequently the network is not capable of giving valid predictions. Receiving incorrect turbulence quantities, the simulation diverges very soon. How to achieve even better stability of the neural network-based turbulence models remains an important question to be further investigated.

\subsection{Scale of the influence region}
\begin{figure}[!htb]
\centering
\includegraphics[width=0.7\textwidth]{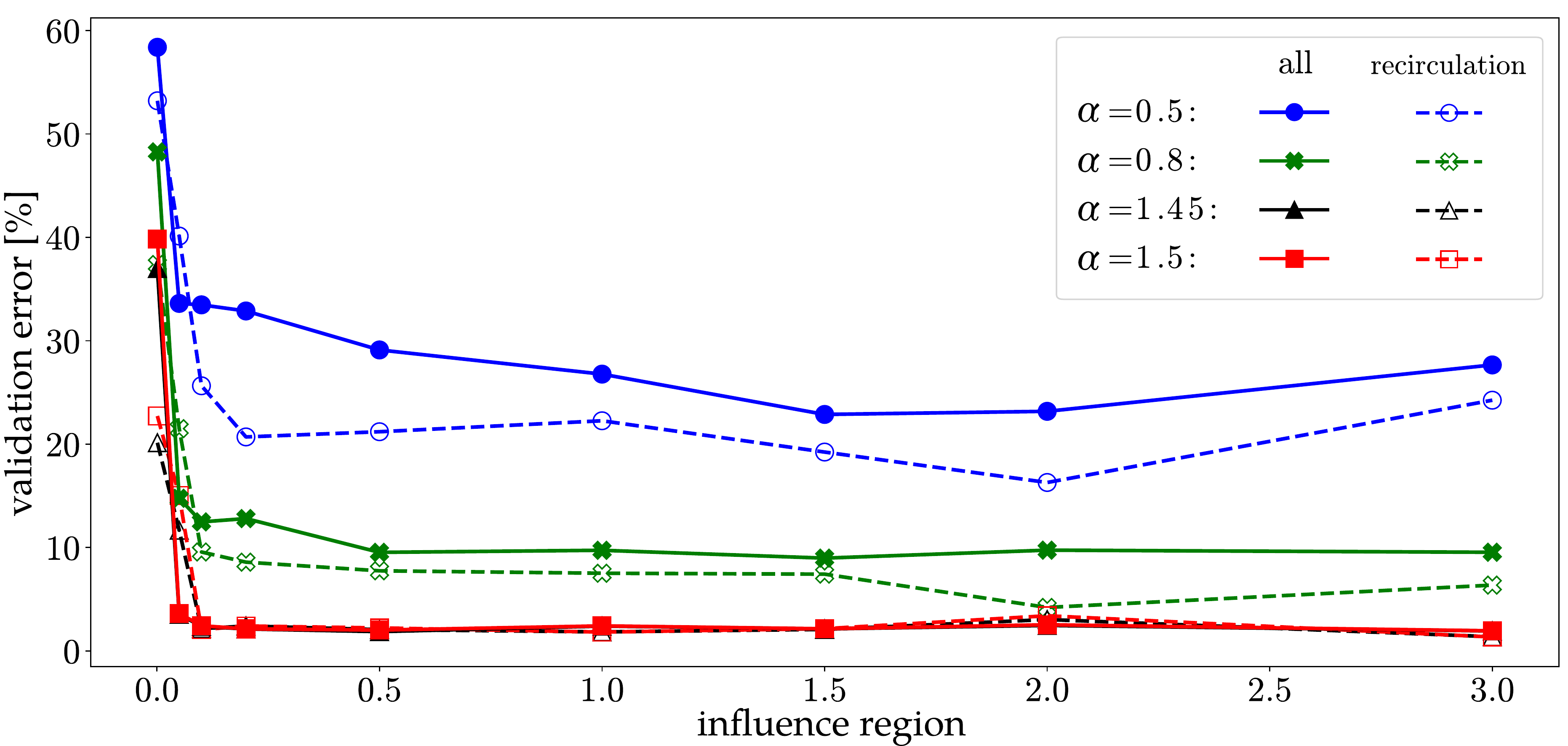}
\caption{Parametric study on the influence region. $x$-axis is the factor with regards to the baseline choice of the major and minor axes of the influence region as provided in Eq.~\eqref{influence_region}. The simulation is performed on $C_I=0, 0.05, 0.1, 0.2, 0.5, 1, 1.5, 2.0, 3.0$. A value of $0$ refers to a local model in which only the mean flow properties of the point of interest itself are regarded as the inputs of the model. The performance of models in the entire flow domain as well as in the recirculation region with various scales of the influence region is evaluated on two extrapolation flow cases $\alpha=0.5$ and $\alpha=0.8$. The recirculation region is marked in (red) shade in \autoref{fig:influence_region_geo} (lower left panel).}
\label{fig:influence_region}
\end{figure}

The parametric study on the scale of the influence region shows a rapidly diminishing marginal benefit of using a large influence region in our case. In \autoref{fig:influence_region}, the validation error is plotted against the coefficient describing the relative size of the influence region as compared to the baseline. This coefficient follows $C_I=\frac{l_1}{l_{1_0}}$, in which ${l_1}$ is the test semi-major axis of the influence region, and $l_{1_0}$ is the original semi-major axis of the influence region based on Eq.~\eqref{influence_region}. The sem-minor axis $l_2$ is scaled by the same constant $C_I$. Therefore, the ratio between the area of the test and baseline influence region is $I/I_0=C_I^2$.

The validation error shows an overall declining tendency with regards to the enlarging influence region. However, there are clearly two stages that can be observed. The first interval is from $C_I=0$ to around $C_I=0.1$, in which the improvement of performance is the most evident. In the second interval ($C_I>0.1$), the performance stays roughly at the same level. 

\begin{figure}[!htb]
\centering
\includegraphics[width=0.75\textwidth]{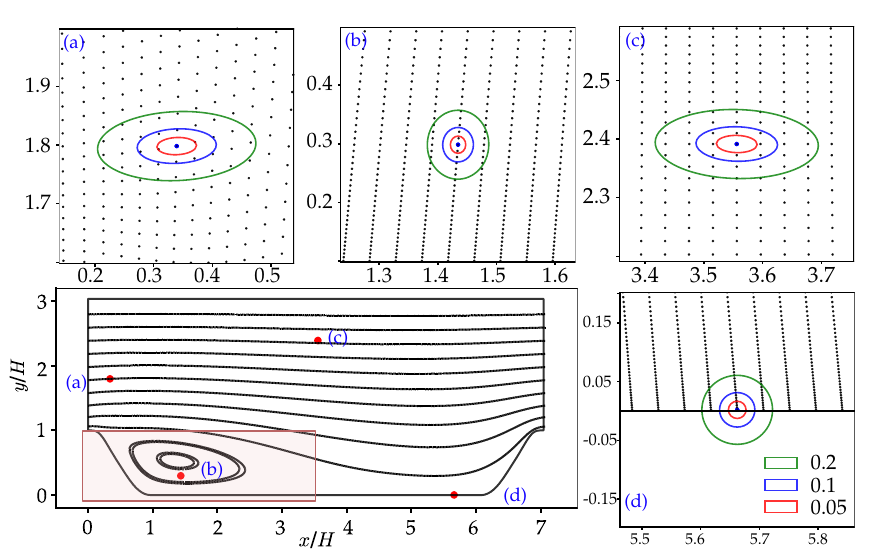}
\caption{Visualization of stencils under various sizes of influence region for flow case $\alpha=0.5$. Four points located at the inlet, the recirculation region, the main stream, and near the bottom are inspected.}
\label{fig:influence_region_geo}
\end{figure}

As indicated in \autoref{fig:influence_region_geo}, when coefficient $C_I$ reaches $0.1$, the stencil in most parts of the flow domain contains neighbouring points both along and orthogonal to the streamwise direction. By including these critical points, the network performance is greatly enhanced. Influence regions larger than this can be considered as containing the most important set of information for predicting turbulence quantities. It is noticed that in the recirculation region, $C_I=0.2$ can also be filed under the first interval. Because in the near-wall region, only until $C_I=0.2$ are the neighbouring points along the main flow direction included.

In the second interval, by including points in a larger neighbourhood, the prediction capacity of the network increases only by a small amount. There is possibly even a decline of performance in flow case $\alpha=0.5$ when the influence region increases from $C_I=2.0$ to $C_I=3.0$. This may be due to the fact that noise consisting of points too far away is introduced and undermines the prediction ability. 
\subsection{Flow physics learned by the neural network}
\begin{figure}[!htb]
\centering
\includegraphics[width=0.6\textwidth]{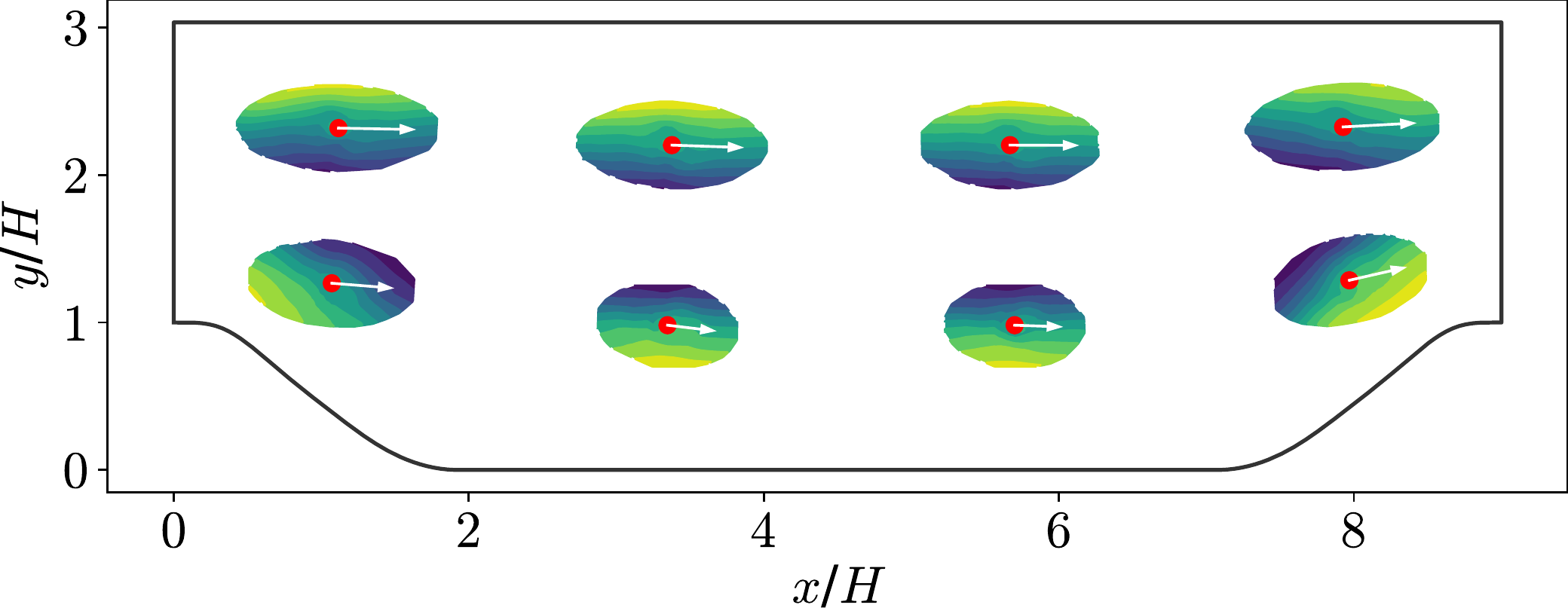}
\caption{Contour plot of embedded weights $G_{i1}$ inside stencils at various locations indexed by $i$. Lighter colors in this figure indicate larger values of the embedded weights and potentially larger influence. The arrows indicate the direction and the magnitude of the local velocity.}
\label{fig:embedding_G1}
\end{figure}

The neural network model has strong boundary dependence as shown in the visualization of one of the embedding functions. We set $m'=1$ for a better interpretation of the network. In this way, only the first embedding function is multiplied twice in the linear transformation step and it is attached with more importance than other embedding functions. The embedded weights $G_{i1}$ defined by the embedding function $\phi_1$ where $i$ denotes the index of spatial points is visualized in \autoref{fig:embedding_G1}. Across stencils at different locations, the contours of the weights showed a consistent pattern. The gradient of the embedded weights is well aligned with the wall-normal direction, which is especially evident for stencils near the periodic hills. The pattern of weights agreed well with the physics that the flow inside a periodic hill channel is dominated by the wall.

\section{Conclusion}\label{conclusion}
In this work, we apply vector-cloud neural network~\cite{zhou2021learning} to address the long-standing turbulence modelling problem. It is a network architecture incorporating the nonlocality of turbulence quantities and guaranteeing the invariance properties of turbulent flows. This work is a proof of concept for the application of VCNN to turbulence modelling in a couple setting and pave the way for the framework to predict Reynolds stress tensor anisotropy.

The neural network-based turbulence model was embedded into the RANS solver and has demonstrated good stability in the coupled setup. We tested the stability and robustness of the coupled solver by applying it at different stages (e.g., $500^\textrm{th}$ iteration, $100^\textrm{th}$ iteration) of the simulation. The neural network coupled solver could bring the simulation to a convergence (with good accuracy) starting from an early stage ($500^\textrm{th}$ iteration) as shown in our testing cases. The stability of neural network models is usually not guaranteed in physical simulations. Thus, our numerical test results demonstrate the promising future of VCNN in other application scenarios.

When adapting the network from laminar to turbulent flows, the wide distribution of turbulence quantities $k$ and $\varepsilon$ brings about challenges for machine learning. We apply a transformation to each of the inputs and outputs. It is shown in numerical tests that the training efficiency has largely improved after transformation. Besides, based on the parametric study on the scale of the influence region, there is a diminishing marginal utility when increasing the influence region. Whether this discovery is universal to other flow cases requires further investigation. Finally, the physics learnt by the network showed a strong boundary relevance which was consistent with the expected flow physics for channel flows. It suggested that the vector-cloud neural network was able to learn the correct nonlocal physics from the provided data.

\end{document}